\documentclass[pre,twocolumn,preprintnumbers,amsmath,amssymb,longbibliography]{revtex4}
\usepackage{amssymb,amsmath,amssymb,amsthm,graphicx}
\usepackage{amsmath,amssymb}
\usepackage{float}
\usepackage{graphicx}
\usepackage{psfrag}
\usepackage{color}
\usepackage{dcolumn}
\usepackage{bm}
\usepackage[normalem]{ulem}
\usepackage[absolute]{textpos}

\def\beq{\begin{equation}}
\def\eeq{\end{equation}}
\def\bea{\begin{eqnarray}}
\def\eea{\end{eqnarray}}
\newcommand{\stkout}[1]{\ifmmode\text{\sout{\ensuremath{#1}}}\else\sout{#1}\fi}

\begin{document}
 
\title{Rough or crumpled: Phases in kinetic growth with surface relaxation}

\author{Sudip Mukherjee}\email{sudip.bat@gmail.com}
\affiliation{Barasat Government College,
10, KNC Road, Gupta Colony, Barasat, Kolkata 700124,
West Bengal, India}
\author{Abhik Basu}\email{abhik.123@gmail.com,abhik.basu@saha.ac.in}
\affiliation{Theory Division, Saha Institute of
Nuclear Physics, 1/AF Bidhannagar, Calcutta 700064, West Bengal, India}

\begin{abstract}
We show that generic  kinetic growth processes with surface relaxations can exhibit a hitherto unexplored crumpled phase with short-range orientational order at dimensions $d<4$. A sufficiently strong  spatially non-local part of the chemical potential associated with the particle current above a threshold in the system can trigger this crumpling. The system can also be in a perturbatively accessible rough phase with long range orientational order but short range positional order at $d<4$ with known scaling exponents. Intriguingly, in $d>4$ we argue that there is no crumpling transition; instead, there is a roughening transition from a smooth to a rough phase for large enough non-local particle chemical potential. Experimental and theoretical implications of these results are discussed.
\end{abstract}

\maketitle



{{Nonequilibrium fluctuations are far more difficult to characterise theoretically due to the combined effects of the nonlinear interactions and the lack of an Fluctuation-Dissipation-Theorem~\cite{chaikin}. The absence of a general theoretical framework for nonequilibrium systems has prompted physicists to work on simple nonequilibrium models, which are amenable to well-controlled calculations, but still produce novel results with wider ramifications. A prominent example is the Kardar-Parisi-Zhang equation (KPZ) equation, originally proposed as a simple nonlinear model of nonconserved surface growth without any overhangs~\cite{kpz,stanley}, that is reinvented as paradigmatic nonequilibrium model undergoing nonequilibrium {\em roughening transition} between a {\em smooth} and a perturbatively inaccessible {\em rough} phase at dimensions $d>2$~\cite{kpz,stanley,kpz-rough}. 
A conserved version of the KPZ (CKPZ) equation driven by a conserved noise~\cite{ckpz-basic} has been proposed that unsurprisingly belongs to a different universality class with no roughening transition at all. A subsequent generalisation of the CKPZ equation having a non-local part of the chemical potential $\mu$ associated with the particle current $\bf J$~\cite{mike} reveals a more complex phase behaviour including new growth phases in a parameter regime for $d>1$. A dynamical model that is closely related to the CKPZ equation is a nonlinear model, originally proposed for ideal Molecular Beam Epitaxy~\cite{das-sarma,stanley} (hereafter the Lai-Das Sarma or LDS equation) { and also subsequently used as a model for tumor growth~\cite{tumor}}. It has the same conservation law structure as the CKPZ equation~\cite{ckpz-basic}, but is driven by a nonconserved noise~\cite{stanley,das-sarma}. Due to the difference in the noise statistics, the CKPZ and LDS equations belong to two different universality classes. 
This naturally begs the question whether a generalisation of the LDS equation similar to the generalised CKPZ equation shows a similar transition, or something else, or nothing at all.}} 

In this Letter, we investigate the universal scaling in a generalised LDS model with $\mu$ having both local and non-local parts, analogous to the generalised CKPZ equation~\cite{mike}. This reduces to the LDS equation~\cite{das-sarma}, when the non-local part of $\mu$ vanishes. We show that for weaker non-local $\mu$, the model belongs to the LDS universality class: At $d<4$, it only has a rough phase whose exponents can be obtained in systematic renormalised perturbation expansions; the scaling exponents are identical to those obtained in Ref.~\cite{das-sarma}. For larger non-local $\mu$, the surface however crumples at $d<4$, as soon as the system size exceeds a finite threshold, {with a concomitant loss of orientational long range order (LRO)}. This forms a hitherto unstudied nonequilibrium analogue of   membrane crumpling, an intriguing phenomenon that is well-studied in the statistical mechanics of membranes in thermal equilibrium~\cite{tethered,david-guitter,john-tethered,john-tethered1}, but not in systems out of equilibrium.  For $d>4$, the model shows a {\em roughening transition} from a smooth to rough surfaces, controlled by the non-local part of $\mu$, with the attendant loss of positional LRO.

We now derive these results. We start with the equation of motion 
\begin{equation}
 \frac{\partial h}{\partial t} = -\nu\nabla^4 h - \frac{\lambda}{2}\nabla^2 ({\boldsymbol \nabla} h)^2 -\lambda_1 {\boldsymbol\nabla}\cdot [({\boldsymbol\nabla} h)\nabla^2 h]+\eta,\label{model-eq}
\end{equation}
where $\nu>0$ is a damping coefficient, $\eta({\bf x},t)$ is a Gaussian noise, and $\lambda,\,\lambda_1$ are the nonlinear coupling constants whose signs are arbitrary; $\lambda_1=0$ gives the LDS equation~\cite{das-sarma}. 
Noise $\eta({\bf x},t)$ is a zero-mean, Gaussian white noise with a variance
\begin{equation}
 \langle \eta({\bf x},t)\eta (0,0)\rangle = 2D\delta ({\bf x})\delta(t).\label{noise-vari}
\end{equation}
Equation~(\ref{model-eq}) has the form $\partial h/\partial t= - {\boldsymbol\nabla}\cdot {\bf J}+ \eta$, where 
\begin{equation}
 {\bf J}= \nu {\boldsymbol\nabla}\nabla^2 h + \frac{\lambda}{2}{\boldsymbol\nabla} ({\boldsymbol \nabla} h)^2 
 +\lambda_1({\boldsymbol\nabla} h)\nabla^2 h.\label{curr-model}
\end{equation}
The first term of the rhs of (\ref{model-eq}), or equivalently, of (\ref{curr-model}), is a ``curvature''-dependent part of the particle current at the linear order in $h$~\cite{das-sarma}. 
It contributes to (\ref{curr-model}) only if there is a gradient in the local mean curvature. Further, the two nonlinear terms with coefficients $\lambda$ and $\lambda_1$ are the symmetry-permitted lowest order nonlinear terms, which are of nonequilibrium origin, and hence cannot be obtained from a free energy. Writing
${\bf J}=-{\boldsymbol\nabla}\mu$ we get
\begin{equation}
\mu=-\nu \nabla^2 h -\frac{\lambda}{2} ({\boldsymbol \nabla} h)^2 - \lambda_1 \nabla^{-2}{\boldsymbol\nabla}\cdot [({\boldsymbol\nabla} h)\nabla^2 h],
\end{equation}
giving $\nabla^2\mu=-{\boldsymbol\nabla}\cdot {\bf J}$.
Thus, the $\lambda_1$-term acts as a {\em non-local} chemical potential, as opposed to the local chemical potential term with coefficient $\lambda$. 
{ { The $\lambda_1$-term, originally introduced in Ref.~\cite{mike} to generalise the CKPZ equation, makes a contribution to $\bf J$ in (\ref{curr-model}) that is bilinear in the local {\em curvature} and {\em tilt}. 
In fact, its effect can been interpreted as ``blind geodesic jumping'', in which particles move a small fixed geodesic distance along
the surface in a random direction~\cite{mike}. As illustrated in Ref.~\cite{mike}, this potentially leads to a current contribution as in the $\lambda_1$-term.}}  This $\lambda_1$-term is responsible for a roughening transition in the generalised CKPZ equation, absent in the original the CKPZ equation, where $\lambda_1=0$ identically~\cite{mike}.  



Universality of the statistical steady states of (\ref{model-eq}) are described by the universal spatio-temporal scaling of the time-dependent correlation function $C$ of $h$:
\begin{equation}
 C(r,t)\equiv \langle [h({\bf x},t)-h(0,0)]^2\rangle \sim r^{2\chi}f_h({r^z}/{t}),\label{corr-h}
\end{equation}
where, $r=|{\bf x}|$, $\chi_h$ and $z$, respectively, are the roughness and dynamic exponents. 

 Renormalisation group (RG) analysis on the LDS equation shows that  fluctuations modify the linear theory values of the scaling exponents at $d<4$; it gives a rough phase  with $\chi=\epsilon/3$ and $z=4-\epsilon/3$ to ${\cal O}(\epsilon)$ where $\epsilon\equiv 4-d$~\cite{das-sarma}. At $d>4$, fluctuation-corrections are irrelevant and the linear theory values of the scaling exponents are obtained with $z=4$ and $\chi=(4-d)/2$. 


{ {Is the $\lambda_1$-term a relevant perturbation on the LDS equation? It has the same number of fields and gradients as the $\lambda$-term, and hence, both the terms are expected to be na\"ively equally relevant in the RG sense. The simplest way to ascertain that is by checking their scaling dimensions: Rescale ${\bf x}\rightarrow b {\bf x},\, t\rightarrow b^z t,\, h\rightarrow b^{\chi}h$ with the linear theory values of $\chi=2-d/2$ and $z=4$, such that $\nu$ and $D$ {\em do not} scale. Under this rescaling, both $(\lambda,\,\lambda_1)\rightarrow b^{4-d}(\lambda,\,\lambda_1)$,  showing their {\em equal relevance} in the RG sense with 4 as their common critical dimension. In fact, there are no symmetry arguments to discard the $\lambda_1$-term, since both of them have the same symmetry being invariant under $h\rightarrow h+const$. This naturally calls for inclusion of the $\lambda_1$-term in a complete RG study, that will allow us to systematically explore the ``new steady states'' due to a non-zero $\lambda_1$, not found in the LDS equation. }} What is the nature of these new steady states?  The nonlinear terms preclude exact enumeration of the scaling exponents. We therefore take a perturbative approach. Conservation law and the invariance under a constant shift of the base plane ensure that the fluctuations of $h$ are long lived; the life-time of the fluctuations of size $q^{-1}$ diverges as wavevector $q\rightarrow 0$. As a result, perturbative corrections to the model parameters diverge in the infra-red limit. 
These divergences are systematically handled within the dynamic RG framework~\cite{stanley,fns,halpin,janssen}. The RG procedure is conveniently implemented  by using a path integral description, equivalent to and constructed from Eq.~(\ref{model-eq}) together with (\ref{noise-vari}), in terms of the field $h({\bf x},t)$, and its dynamic conjugate field $\hat h({\bf x},t)$~\cite{janssen}.

We perform the one-loop Wilson momentum shell dynamic RG procedure~\cite{fns,stanley,halpin}. Since the $\lambda_1$-term in Eq.~(\ref{model-eq}) can be written as a combination of $\nabla^2 ({\boldsymbol\nabla} h)^2$ and $\nabla_i\nabla_j (\nabla_i h \,\nabla_j h)$, fluctuations do not generate a $\nabla^2 h$-term in Eq.~(\ref{model-eq}). The resulting fluctuation-corrections to the model parameters are represented by the one-loop Feynman diagrams; see Supplemental Material (SM)~\cite{sm}. 
There are {\em no} relevant fluctuation-corrections to the noise amplitude $D$, which is an {\em exact} statement. This is due to the fact that the noise in (\ref{model-eq}) is non-conserved, whereas the equation of motion (\ref{model-eq}) has the form of a conservation law. This means the corrections to $D$ is at ${\cal O}(q^2)$, or higher, whereas the bare noise amplitude is ${\cal O}(q^0)$, ruling out any relevant (in the RG sense) corrections to it. Furthermore, there are no relevant one-loop corrections to $\lambda$ or $\lambda_1$~\cite{janssen-prl}. There are however diverging one-loop corrections to $\nu$. By dimensional analysis we identify two dimensionless effective coupling constants
\begin{equation}
 g= \frac{\lambda^2 D}{\nu^{3}}K_d\Lambda^{d-4},\;\gamma=\frac{\lambda_1}{\lambda}.
\end{equation}
Here, $K_d$ is the surface area of a $d$-dimensional hypersphere of unit radius.
Further $g$ has a critical dimension $d_c=4$ as in Ref.~\cite{das-sarma}. 
By following the standard steps of RG outlined in ~\cite{stanley}, we obtain the following differential RG recursion relation to the linear order in $\epsilon\equiv 4-d$:
\begin{eqnarray}
  \frac{d\nu}{dl}&=&\nu[z-4+g\Delta(\gamma)],\label{nu-rec}\\
 \frac{d(\lambda,\,\lambda_1)}{dl}&=&(\lambda,\,\lambda_1)[\chi+z-4],\label{l-rec}\\
 \frac{dD}{dl}&=&D[z-d-2\chi],\\
 \frac{dg}{dl}&=&g[\epsilon-3g\Delta(\gamma)],\label{flow-G}
\end{eqnarray}
where, $\Delta(\gamma)=\left(-\frac{3\gamma^2}{8}+\gamma+\frac{1}{8}\right)$, $b\equiv\exp(l)$ is a length-scale; $\xi=b^\chi$, with $\chi$ being the roughness exponent. With $\lambda_1=0$, (\ref{nu-rec}) and (\ref{flow-G}) reduce to those in the LDS equation~\cite{das-sarma}. The surface is rough or smooth, if $\chi>0$ or $<0$. 
%
Since $\lambda$ and $\lambda_1$ can have either sign independently of each other, the dimensionless ratio $\gamma$ can also be positive or negative. The nonrenormalisation of both $\lambda$ and $\lambda_1$ at the one-loop order means their ratio $\gamma\equiv  \lambda_1/\lambda$ too does not renormalise, and is {\em marginal}: $d\gamma/dl=0$. 


Depending upon the signs of $\epsilon$ and $\Delta$, there are four distinct cases obtained from
 $dg/dl=0$ at the RG fixed point. We first focus on the cases with $\epsilon>0$ (i.e., $d<4$), which includes the physically relevant dimension 2.
 


(i) If $\epsilon>0$ ($d<4$) and $\Delta>0$, then 
\begin{equation}
 g=0,\; g=\frac{\epsilon}{3\Delta}=g^*(\gamma)  \label{g-fixed}
\end{equation}
 as the fixed points of $g$.
Out of these two fixed points,
$g= g^*(\gamma)$ is stable. It is, in fact, a {\em fixed line} in the $g-\gamma$ plane, parametrised by $\gamma$. On this fixed line, by using $d\nu/dl=0$, we get
%
\begin{eqnarray}
 z&=& 4-g\Delta = 4-\frac{\epsilon}{3},\;
 \chi= \frac{z-d}{2}=\frac{\epsilon}{3}.\label{expo2}
\end{eqnarray}
These are identical to their counterparts in the LDS equation (i.e., $\gamma=0$)~\cite{das-sarma}, and form the LDS universality class. Thus, at this stable fixed point, valid when $\Delta>0$ together with $\epsilon>0$, model equation (\ref{model-eq}) belongs to the LDS universality class~\cite{das-sarma} for all $\gamma$.

(ii) Now consider $\epsilon>0$ ($d<4$) and $\Delta<0$. This gives
\begin{equation}
 \frac{dg}{dl}=g\left[\epsilon+ 3g|\Delta|\right].\label{flow-eq-2}
\end{equation}
In this case, $g=0$ is the only fixed point, which is unstable. We argue below that this represents crumpling of the surface.

\begin{widetext}

\begin{figure}[htb]
 \includegraphics[width=5cm,height=4.8cm]{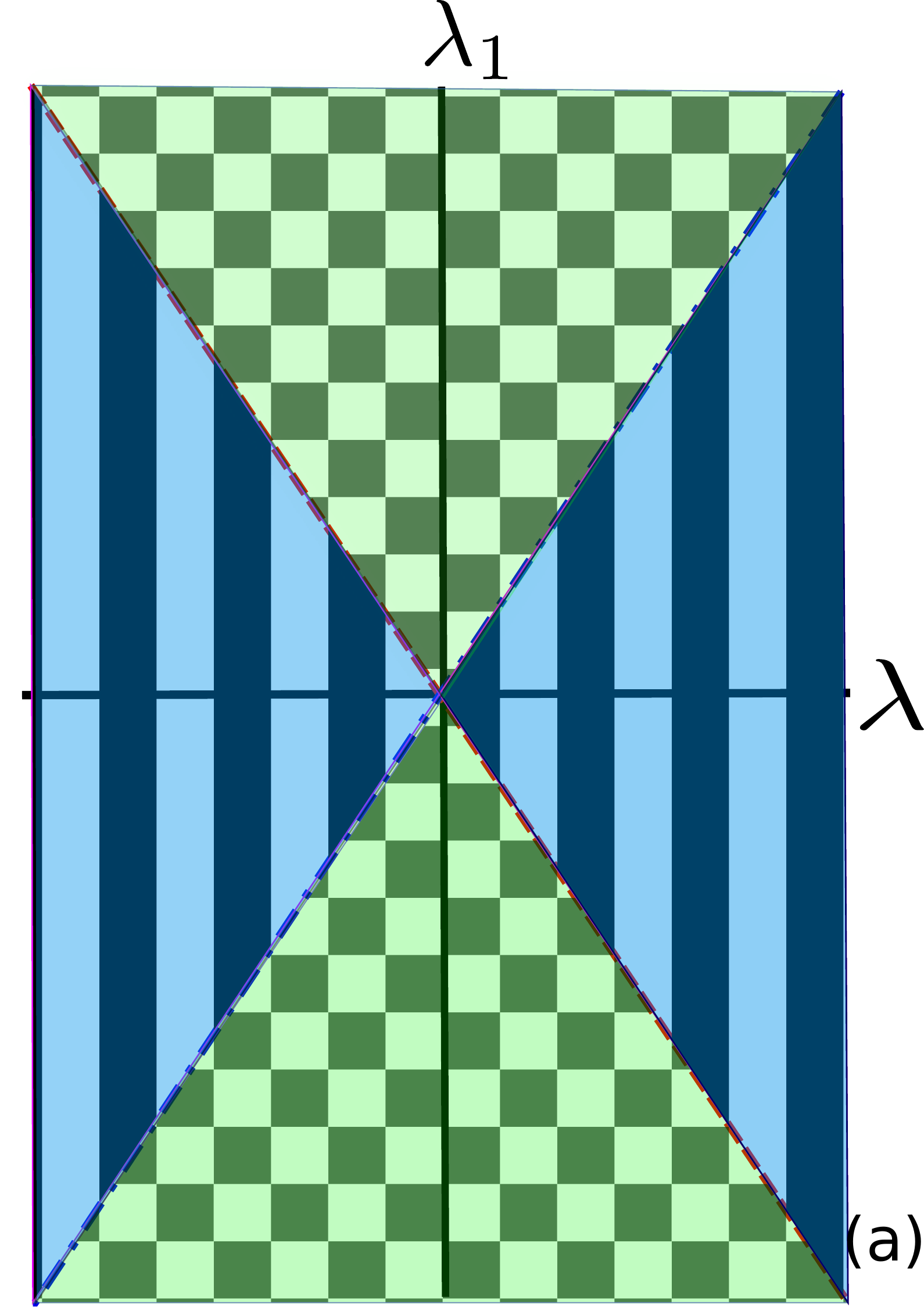}\hfill \includegraphics[width=5.5cm,height=4.8cm]{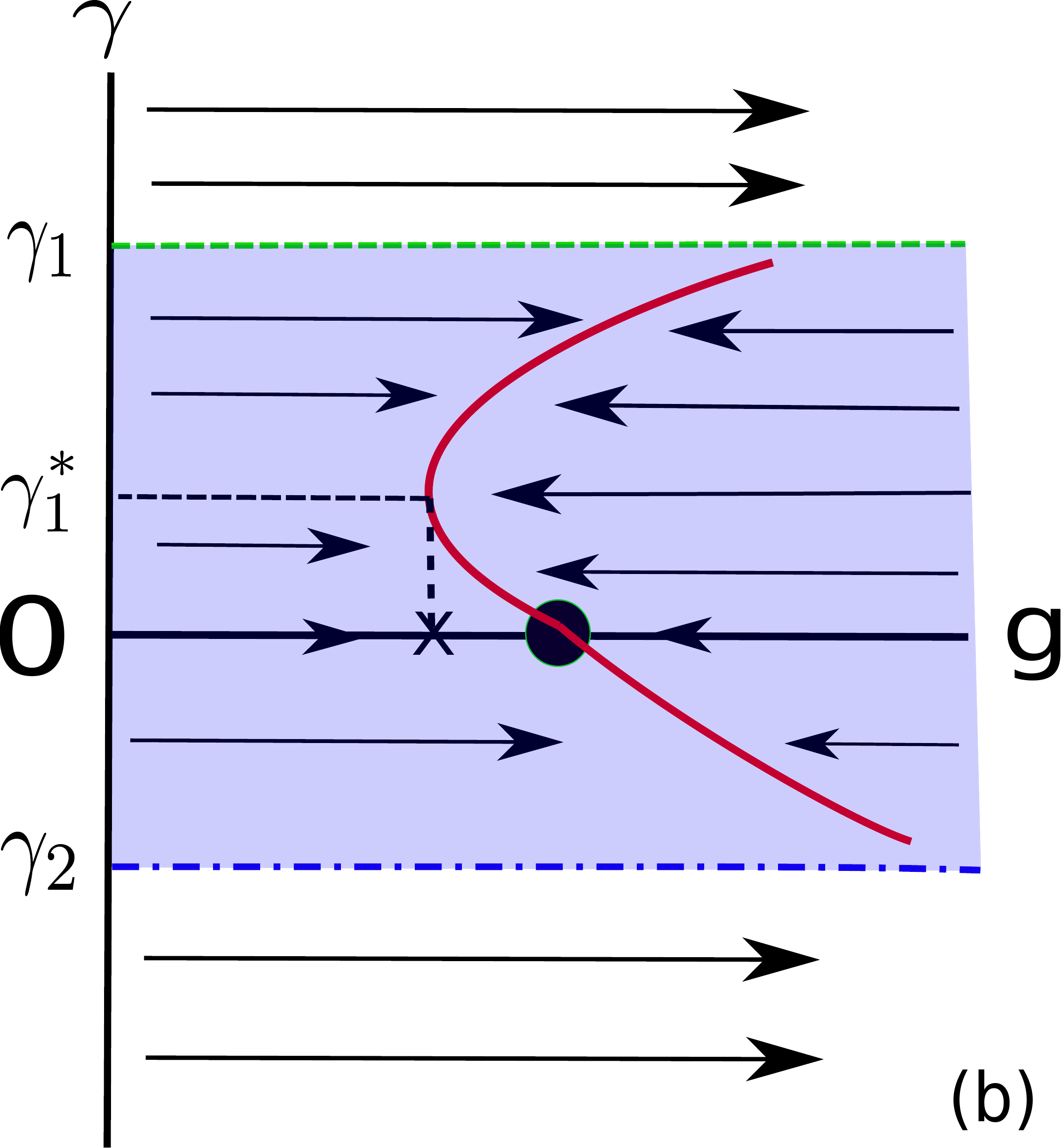} \hfill \includegraphics[width=5cm,height=4.8cm]{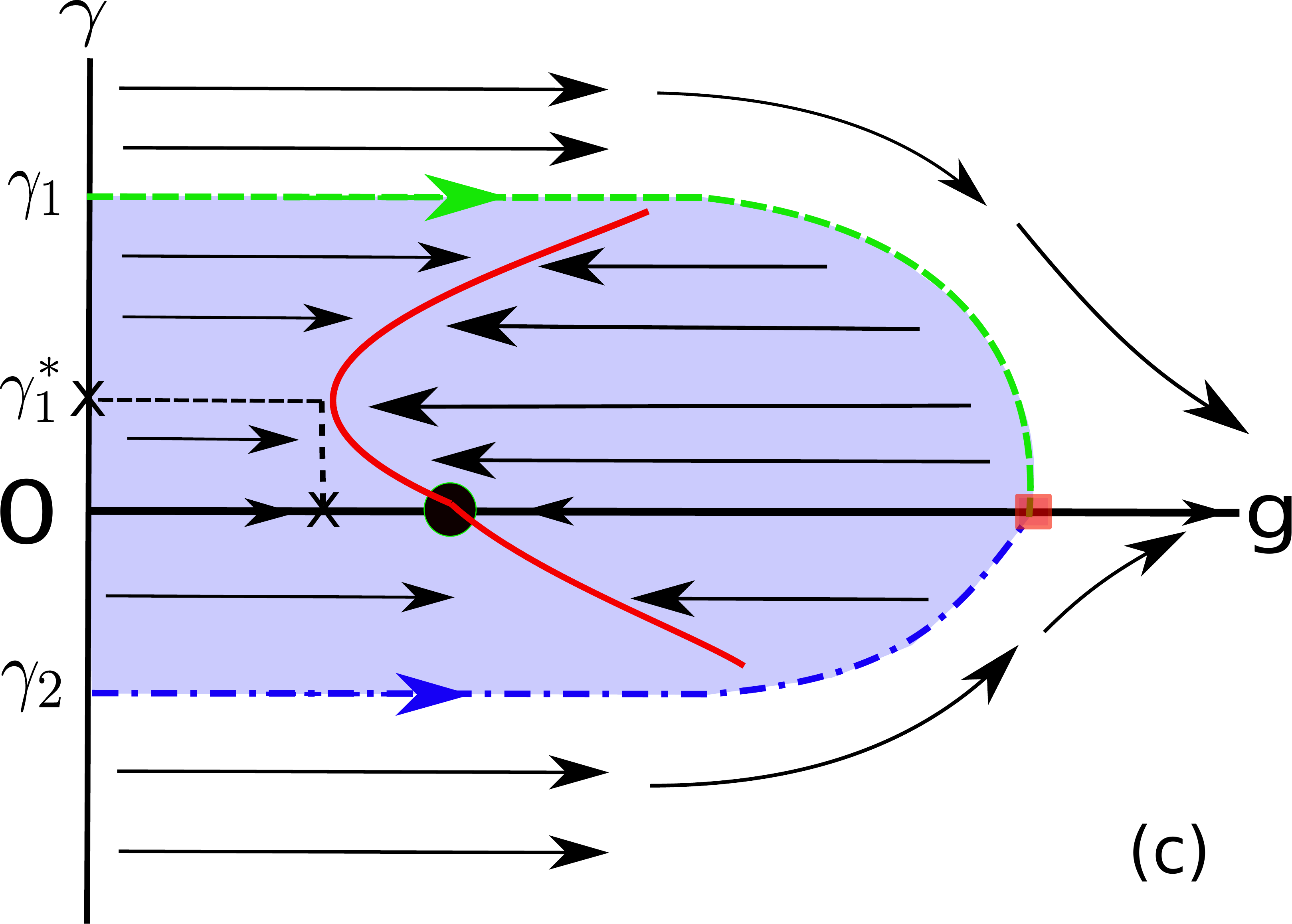}
 \caption{(colour online)(a) Schematic phase diagram in the $\lambda-\lambda_1$ plane ($d<4$), showing the regions corresponding to a rough (stripes) and a crumpled (checkerboard) phase. (b) Schematic phase diagram and the RG flow lines in the $g-\gamma$ plane ($d<4$). The small filled circle on the $g$-axis is the original LDS fixed point $g=8\epsilon/3$. The red curved line is the {\em fixed line} $g=\epsilon/(3\Delta(\gamma))$. (c) Conjectured ``Occam's razor'' global RG flows in the $g-\gamma$-plane ($d<4$). Arrows indicate the flow directions. The green and blue broken lines give the boundary between the stable (rough) and the crumpled phases. The small filled red square on the $g$-axis is the putative unstable fixed point not accessible in our perturbative RG (see text). 
 }\label{lam-lam1}
\end{figure}

\end{widetext}

We note that the above two cases are distinguished by the sign of $\Delta$. Naturally, $\Delta=0$ gives the separatrix between the two kinds of behaviour, giving two solutions $\gamma_\pm$  for $\gamma$, or $\lambda_1$ in terms of $\lambda$: { {
\begin{eqnarray}
 \gamma_\pm &=&  \frac{1}{3}\bigg[4\pm\sqrt{19}\bigg]>(<)0,\label{gamval1}\\
 \lambda_1&=&\frac{\lambda}{3}\bigg[4\pm \sqrt{19}\bigg], \label{lam-bound}
\end{eqnarray}
}} giving the boundaries of stability in the $\lambda-\lambda_1$ plane; see Fig.~\ref{lam-lam1}(a) for a schematic phase diagram and discussions below.
For $\gamma>\gamma_+$, or $\gamma<\gamma_-$, $\Delta<0$, corresponding to case (ii), giving instability. In contrast, when $\gamma$ belongs to the other range, i.e., $\gamma_+>\gamma>\gamma_-$, $\Delta >0$, we get $g=g^*(\gamma)=\epsilon/(3\Delta)$ as the stable fixed point. As noted earlier, the stable case (i) belongs to the LDS universality class~\cite{das-sarma}. The unstable case (ii) is {\em new}, and is due to the nonlinear term with coefficient $\lambda_1$ in Eq.~(\ref{model-eq}). Since $\gamma$ is {\em marginal} at the one-loop order, we obtain a {\em fixed line} in the $g-\gamma$ plane in the stable case.  In fact from (\ref{g-fixed}),
if $\gamma=0$, unsurprisingly the LDS  fixed point is recovered: $g^*(\gamma=0)=8\epsilon/3$. For $\gamma_{1}>\gamma\geq 0$, we note that initially for very small $\gamma>0$, $g^*(\gamma)$ {\em decreases} as $\gamma$ rises, but eventually starts to rise as $\gamma$ crosses a threshold $\gamma_{+}^*=4/3$, which maximises $\Delta$, and diverges as $\gamma\rightarrow \gamma_{+}$ from below. Thus $g^*$ has a nonmonotonic dependence on $\gamma$ when $\gamma_+>\gamma\geq 0$. We find that $g^*(\gamma=\gamma_{+}^*)=19\epsilon/24$, which is the smallest value of $g^*$ possible. In contrast, in the range $\gamma_{-}<\gamma<0$, $g^*$ rises monotonically as $\gamma$ decreases, eventually diverging as $\gamma\rightarrow\gamma_{-}$ from above. Since $\gamma$ is marginal in the one-loop approximation, within the range $\gamma_{-}<\gamma<\gamma_{+}$ the RG flow lines run parallel to the $g$-axis towards the ``fixed line'' described above. Outside this regions, the flow lines flow parallel to the $g$-axis towards infinite $g$, indicating breakdown of the perturbation theory. See Fig.~\ref{lam-lam1}(b) for a schematic RG flow diagram in the $g-\gamma$ plane.

We can solve Eq.~(\ref{flow-eq-2}) to get 
\begin{equation}
 g(l)=\frac{\epsilon}{3|\Delta|}\frac{A\exp (\epsilon l)}{1-A\exp(\epsilon l)}.\label{g-soln}
\end{equation}
The constant of integration $A$ can be evaluated by using the ``initial'' condition: at $l=0,\,g(0)=g_0$. This gives
 $A=({g_0}/{\epsilon}){3|\Delta| + g_0})$.
If $\Delta \rightarrow 0$ from below, we get
 $A\approx g_0 (\frac{3|\Delta|}{\epsilon})$,
corresponding to
$ g(l)\approx g_0 e^{\epsilon l},$
giving the initial growth of $g(l)$ before the nonlinear effects become important.
For any finite $\Delta<0$, $g(l)$ diverges at a finite $l=l_c$ if
 $1-A\exp(\epsilon l_c)\rightarrow 0_+$,
giving
 $l_c=\frac{1}{\epsilon} \ln A$
as the critical ``RG time'' in which $g(l)$ diverges. 

We now use (\ref{g-soln}) to obtain the scale-dependent renormalised $\nu(l)$. Substituting $g(l)$ in (\ref{nu-rec}),
we find
\begin{equation}
 \nu(l)=\nu_0\exp\bigg[-\frac{\epsilon}{3}\int dl \frac{A \exp(\epsilon l)}{1-A\exp(\epsilon l)}+(z-4)l\bigg],\label{nu-van}
\end{equation}
which clearly shows that $\nu(l)$ vanishes as $l\rightarrow l_c\sim {\cal O}(1)$ from below, or equivalently, as the length-scale dependent $\nu(L)$ vanishes as $L\rightarrow \xi$ from below, where $\xi=a_0 \exp (l_c)$; $a_0$ is a small-scale cutoff. We thus identify $\xi$ as the {\em persistence length}, beyond which the system crumples. Of course, the perturbation theory actually breaks down as soon as $g(l)$ becomes ${\cal O}(1)$. 

We now argue that case (ii) represents a {\em crumpled phase}, quite distinct from a rough surface. A rough surface is characterised by (a) a positive  $\chi>0$, which implies $\langle h^2({\bf x},t)\rangle$ growing with the system size $L$, which in turn implies a lack of positional LRO, together with (b) a finite (i.e., $L$-independent in the limit of large $L$) $\langle ({\boldsymbol\nabla } h)^2\rangle $ meaning {\em orientational LRO}. For instance, in the LDS equation, $\chi=\epsilon/3>0$. Therefore,  $\langle h^2({\bf x},t)\rangle\sim L^{2\epsilon/3}$, growing indefinitely with $L$. On the other hand, $\langle ({\boldsymbol\nabla } h)^2\rangle$ remains $L$-independent. In contrast, a crumpled surface not only lacks positional LRO, it {\em also lacks orientational LRO.} This is reflected in the divergence of $\langle ({\boldsymbol\nabla } h)^2\rangle$ in the limit of large $L$. It is clear that the LDS rough surfaces are {\em not} crumpled, since $\langle ({\boldsymbol\nabla } h)^2\rangle$ remains $L$-independent, which is similar to the equilibrium tethered membranes in their low temperature ($T$) orientationally ordered phase~\cite{tethered}. 
%
 In contrast, finite but sufficiently large two-dimensional (2D) equilibrium fluid or lipid membranes necessarily crumple at any non-zero temperature~\cite{chaikin,peliti}. Thus a rough (but non-crumpled) and a crumpled surface are distinguished by the presence or absence of long-range orientational order. 
While the phenomenon of crumpling should be rather generic, independent of the conditions of equilibrium, a nonequilibrium example of crumpling in a fluctuating surface has not yet been found.

We further note that in the crumpled phase of the present study, the fluctuation-corrections to $\nu$ are {\em negative}, meaning $\nu(l)$ vanishes in a finite RG time $l_c=\ln (\xi/a_0)$; see (\ref{nu-van}) above. 
%
%
Indeed, at this stage we can make a formal connection with the crumpling of equilibrium statistical mechanics of membranes. Notice that 
$D/\nu_\text{eff}$ has the direct correspondence with $T/\kappa_\text{eff}$ of the membrane, where $\kappa_\text{eff}$ is an ``effective'' bend modulus of the membrane, and $\nu_\text{eff}$ is the fluctuation-corrected, effective $\nu$ of the surface, calculated in the bare perturbation theory. In the case of fluid~\cite{peliti,tirtha-mem1} or asymmetric tethered~\cite{john-tethered,john-tethered1} membranes, $\kappa_\text{eff}$ vanishes, and consequently {\em both} {\em both} $\langle h^2({\bf x},t)\rangle$ and $\langle ({\boldsymbol\nabla}h)^2\rangle$ diverge as soon as the membrane size exceeds a finite threshold, a telltale signature of crumpling. Likewise here, $\nu_\text{eff}$ vanishes and along with it {\em both} $\langle h^2({\bf x},t)\rangle$, and $\langle ({\boldsymbol\nabla}h)^2\rangle$ diverge, as soon as the system size exceeds a finite threshold, giving a nonequilibrium crumpling.   Clearly, the surface morphology in the crumpled phase is fundamentally different from that in the rough phase in the LDS equation.


Thus in the $\lambda-\lambda_1$ plane, for $d<4$ (\ref{lam-bound})
gives the boundaries between regions with a rough phase belonging to the LDS universality class, and a crumpled phase; see the schematic phase diagram in Fig.~\ref{lam-lam1}(a). We further show the RG flow lines ($d<4$) in the $g-\gamma$ plane in Fig.~\ref{lam-lam1}(b), which run parallel to the $g$ axis. For $\gamma_- <\gamma<\gamma_+$, flow lines approach a {\em fixed line}, whose each point corresponds to the LDS universality class. Outside this window, the flow lines run to infinity along the $g$-direction, indicating crumpling of the surface.

We have found that $g^*\rightarrow \infty$ as $\gamma\rightarrow \gamma_{+} (\gamma_-)$ from below (above). However, long before $g$ diverges, our perturbation theory breaks down. In fact, it loses validity as soon as $g^*\sim {\cal O}(1)$. While we cannot obviously follow the RG flows all the way to infinity, we can speculate about the nature of the phases in the region of the parameter space where the RG flow lines appear to run away to infinity. For this, we are guided by the general understanding that for large enough noise any system, equilibrium or nonequilibrium, should undergo a phase transition from a low-noise (low-$T$ in equilibrium systems) ``ordered phase'' to a high-noise (high-$T$ in equilibrium systems) ``disordered phase''. Accordingly, the physical expectation in the present model is that for large enough $g$ (which here means large enough noise or $D$, generalization of $T$ in equilibrium), the system at $d<4$ undergoes a phase transition from a low-noise orientationally ordered rough phase to a phase with short-range order (SRO) only (implying no orientational LRO), i.e., the crumpled phase. Thus, upon increasing $g$, the rough phase ($d<4$) should be unstable, and the surface should eventually crumple. This should hold even in the original LDS equation ($\gamma=0$), for sufficiently $g$: we expect an unstable ``crumpling fixed point'' beyond which the RG flow runs away to infinity signalling loss of orientational LRO. Accordingly, there must be an unstable critical point on the $g$-axis, controlling this transition to this putative crumpled phase. If we now consider the full RG flows for a surface  in the two dimensional parameter space ($g,\,\gamma$)
and connect this putative flow with our
flows for small $g$ and $\gamma$ in the simplest possible way (i.e., without introducing any other new fixed points), we are then led to Fig.~\ref{lam-lam1}(c). This is basically an ``Occam's razor''-style argument: Fig.~\ref{lam-lam1}(c) has the simplest flow topology that naturally reduces to the known flow trajectories for small $g,\,\gamma$ (as shown in Fig.~\ref{lam-lam1}(b)). At the same time, it gives the putative global flow lines, allowing for a transition to a presumed crumpled phase.

We now consider the cases with $d>4$ (i.e., $\epsilon <0$), which although are not physically accessible, provide interesting theoretical insights.

(iii) Consider now $\epsilon <0$ ($d>4$) and $\Delta <0$. This gives
\begin{equation}
 \frac{dg}{dl}=g\left[-|\epsilon| + 3g|\Delta|\right],
\end{equation}
giving $g^*=0$ as the stable fixed point for a {\em smooth phase} with positional LRO, and $g^*=|\epsilon|/(3|\Delta(\gamma)|)$ as the unstable fixed line, parametrised by $\gamma$. This is reminiscent of a roughening transition of the KPZ equation~\cite{stanley}. We also note that the fluctuation-corrections to $\nu$ remain finite even for $L\rightarrow \infty$ for $d>4$, which rules out vanishing of $\nu_\text{eff}$, in turn precluding crumpling as above. Instead, we obtain a roughening transition between a smooth phase ($g^*=0$) and a perturbatively inaccessible rough phase, akin to the roughening transition in the KPZ equation at $d>2$~\cite{stanley}.

(iv) Finally, for $\epsilon <0$ ($d>4$) and $\Delta >0$
\begin{equation}
 \frac{dg}{dl}=g\left[-|\epsilon| - 3g\Delta\right], 
\end{equation}
giving $g^*=0$  is the only fixed point, that is stable.

\begin{widetext}
 
\begin{figure}[htb]
\includegraphics[width=5cm,height=4.1cm]{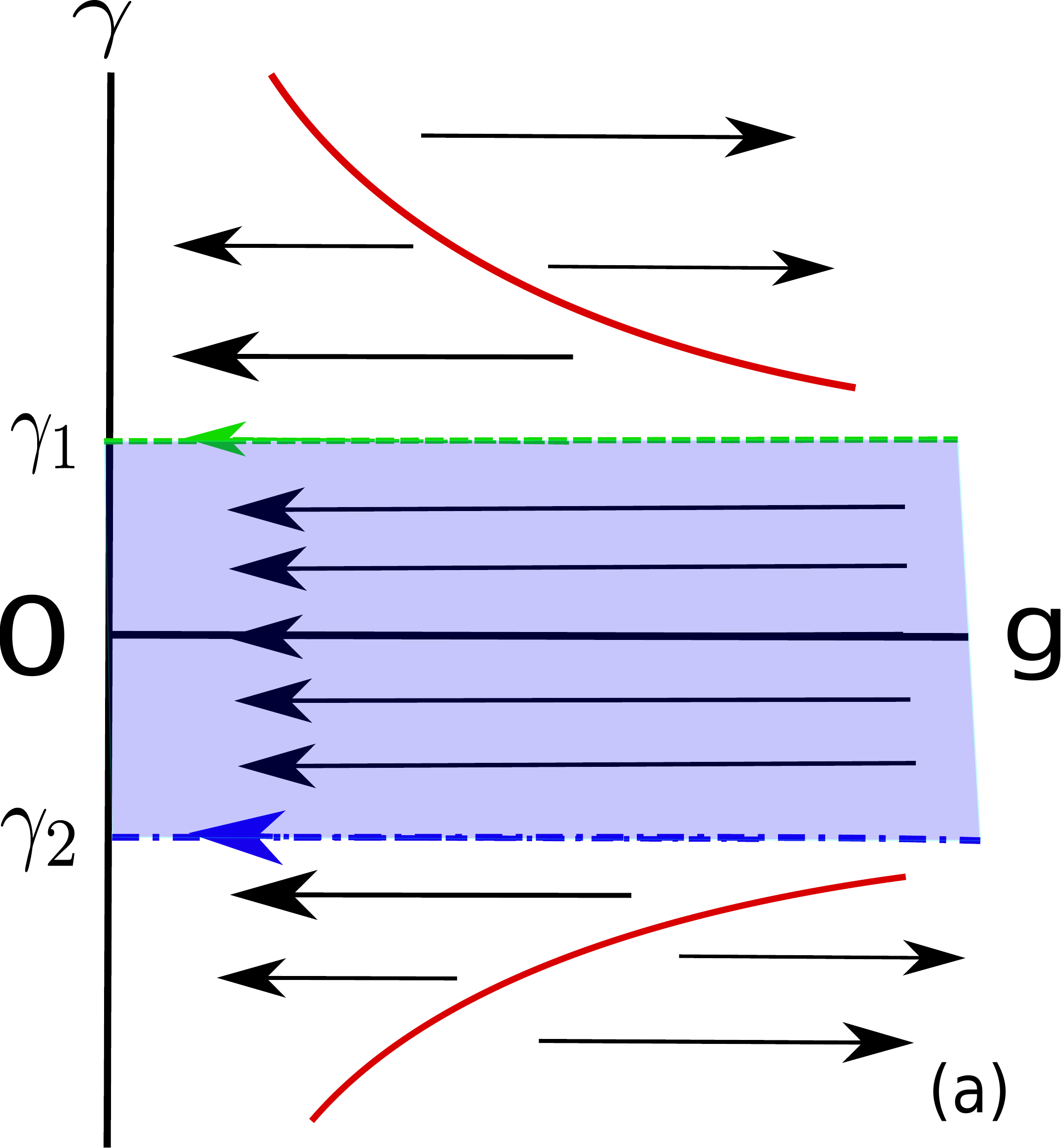}\hfill\includegraphics[width=5cm,height=4.1cm]{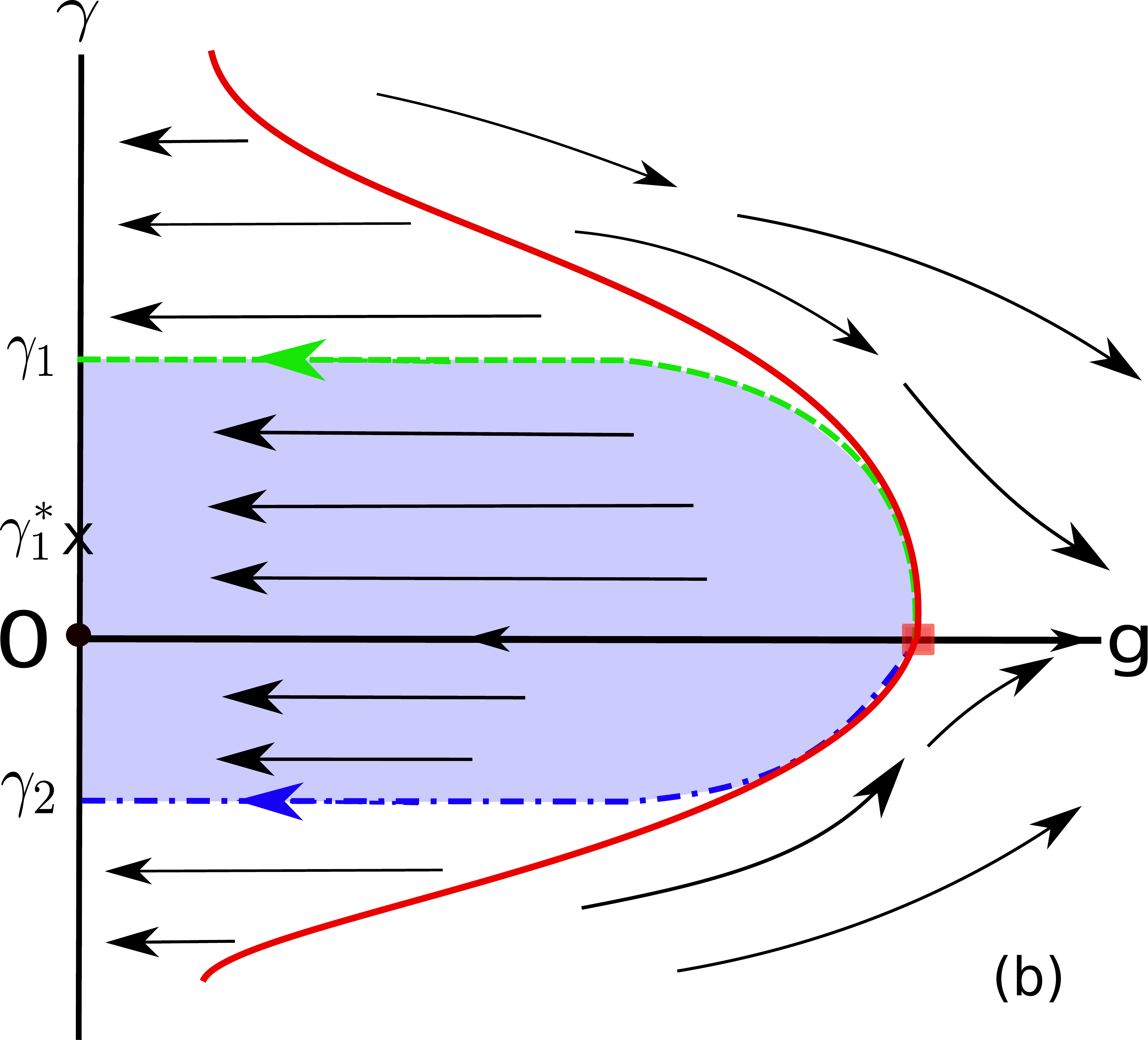}\hfill  \includegraphics[width=5cm,height=3.8cm]{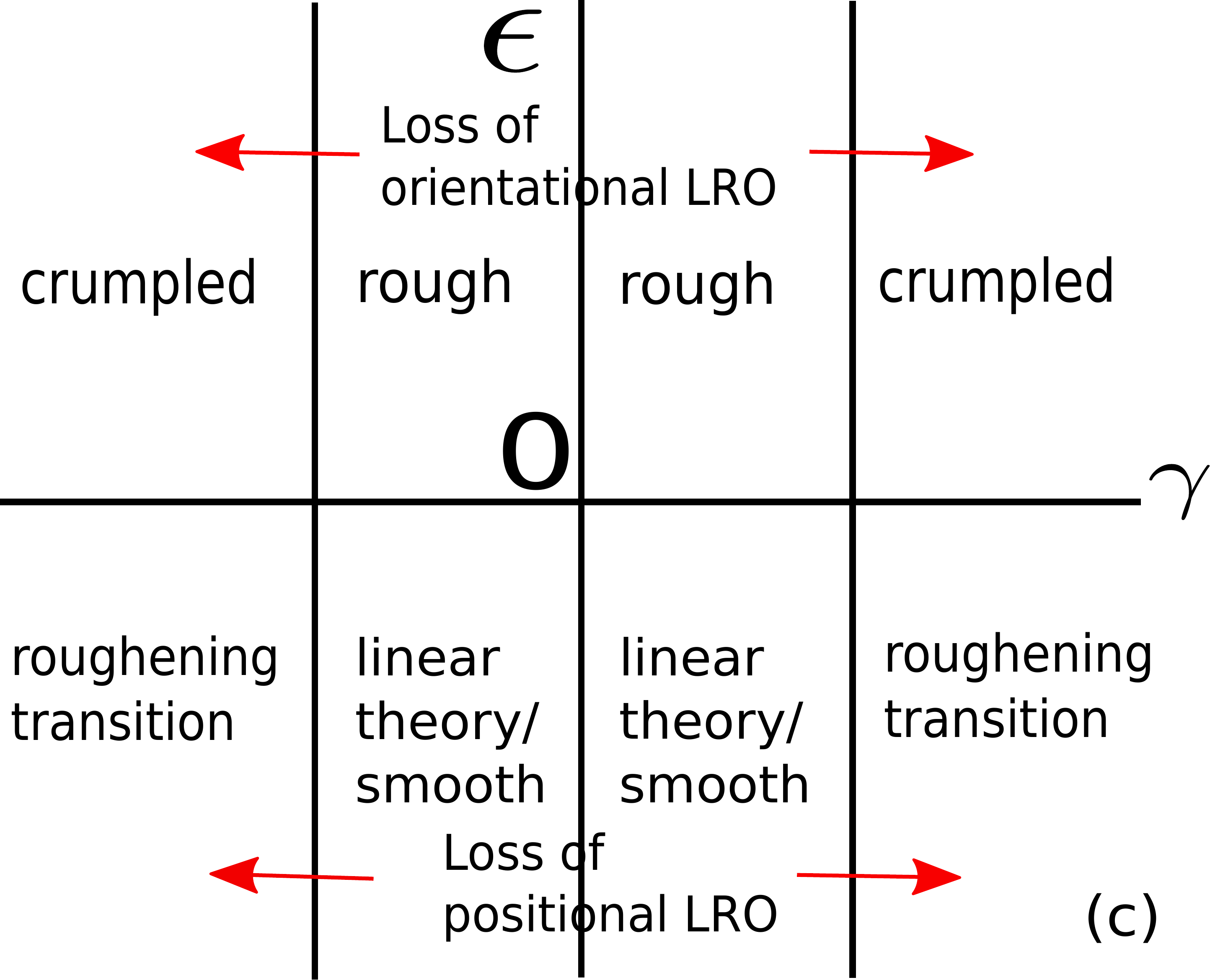}
 \caption{(colour online) (a) Schematic phase diagram and the RG flow lines in the $g-\gamma$ plane ($d>4$). The red curved lines are the {\em unstable fixed lines} given by $g=\epsilon/3|\Delta|,\,\Delta <0$. The shaded region has only a smooth phase; elsewhere there is a roughening transition, (b) Conjectured ``Occam's razor'' global RG flows in the $g-\gamma$-plane ($d>4$). Arrows indicate the flow directions. The red curved lines are the {\em unstable fixed lines} that reduces to $g=\epsilon/3|\Delta|,\,\Delta <0$, in the lowest order perturbation theory. (c) Phases in the $\gamma-\epsilon$ plane. Loss of order across the transitions is shown (see text).}\label{high-d}
\end{figure}

\end{widetext}

The RG flow for cases (iii) and (iv) are shown in Fig.~\ref{high-d}(a). We could again apply an Occam's razor style argument as above. Unlike $d<4$, we now expect loss of positional LRO of the smooth phase upon increasing $g$, resulting into transitions to a rough phase with no positional LRO (but with orientational LRO): we  assume a putative unstable roughening fixed point in the $g$-axis for high enough $g$, and arrive at the putative global RG flow diagram shown in Fig.~\ref{high-d}(b). For $d>4$, Fig.~\ref{lam-lam1}(a) gives the phase boundaries between regions having only smooth phases with no phase transitions and regions with roughening transitions.



For $d>4$, (\ref{lam-bound}) give the boundaries between smooth phase ($\chi<0$) and roughening transition to a perturbatively inaccessible rough phase akin to the rough phase of the KPZ equation for $d>2$.



The four distinct cases depending upon $\gamma$ and $\epsilon$ are schematically shown in Fig.~\ref{high-d}(c).


To summarise, by suitably generalising the LDS equation, we have uncovered an intriguing crumpling instability in $d<4$ in addition to the well-known perturbatively-accessible rough phase which is controlled by the LDS fixed point. The crumpling is controlled by the relative strength of a spatially non-local chemical potential vis-\'a-vis the  corresponding local part. We further show that for $d>4$, instead of a crumpling transition, the system undergoes a roughening transition for a sufficiently large non-local chemical potential, which is akin to the roughening transition of the KPZ equation at $d>2$.  Thus, in a direct analogy with the significance of the KPZ equation 
as a paradigmatic nonequilibrium model with a roughening transition, our work establishes the generalised LDS equation as a simple nonequilibrium model that shows {\em both} crumpling and roughening transitions. We note that in Eq.~(\ref{model-eq}) a $\nabla^2h$-term can in principle be added without violating the conservation law. Such a term remains unaffected by fluctuations. Our theory applies when the coefficient of this term vanishes. Our work can be further extended and complemented by applying various nonperturbative and numerical methods to study the crumpled and rough phases in this model. We conclude with a cautionary remark that in order to distinguish between a rough (having orientational LRO) and a crumpled surface (with SRO), one must measure the variance $\langle ({\boldsymbol \nabla} h)^2\rangle$ of the orientation fluctuations, which for a rough surface is {\em finite} in thermodynamic limit, but diverges for a crumpled surface. 







{\em Acknowledgement:-} The authors thank J. Toner for many useful discussions. S.M. thanks 
the SERB, DST (India) for partial financial support through the TARE scheme [file no.: TAR/2021/000170] (2022). A.B. thanks 
the SERB, DST (India) for partial financial support through the MATRICS scheme [file no.: MTR/2020/000406] (2021).



\pagebreak

\appendix

\section{Action functional}

Model equation~(1) of the main text can be converted to a path integral over configurations given by $h({\bf x},t)$ and its dynamic conjugate $\hat h({\bf x},t)$, which can be used to conveniently calculate the various correlation and vertex functions in the problem. The generating functional is given by
\begin{equation}
 {\cal Z}=\int {\cal D}h {\cal D}\hat h \exp (-S),\label{gen-func}
\end{equation}
where $S$ is the action functional given by

\begin{widetext}

\begin{equation}
 S=\int d^dr dt \bigg[-D\hat h({\bf r},t)\hat h({\bf r},t)\bigg] + \int d^dx dt\,\hat h \bigg[\partial_t h +\nabla^2\{\nu\nabla^2 h + \frac{\lambda}{2}({\boldsymbol\nabla}h)^2\}+\lambda_1 {\boldsymbol\nabla}\cdot\{(\nabla^2 h){\boldsymbol\nabla}h\}\bigg].\label{act-func}
\end{equation}

\end{widetext}
 
 We can now write down the ``bare'' two-point correlation functions in the problem by using (\ref{gen-func}) and the harmonic or Gaussian part of (\ref{act-func}); i.e., by setting $\lambda=0=\lambda_1$. We obtain
 \begin{eqnarray}
  &&\langle |\hat h({\bf q},\omega)|^2\rangle =0,\\
  &&\langle \hat h({\bf -q},-\omega)h({\bf q},\omega)\rangle = \frac{1}{-i\omega +\nu q^2}.\\
  &&\langle |h({\bf q},\omega)|^2\rangle = \frac{2D}{\omega^2+\nu^2 q^4}.
 \end{eqnarray}



\section{Renormalization group analysis }

We outline here the basic steps of the RG calculations performed in the main text.
The RG is done by tracing over the short
wavelength Fourier modes of the fields $h({\bf x},t)$ and $\hat h({\bf x},t)$ (see Refs.~[14-16] of the main text). In particular, we  follow the usual approach of initially restricting the wavevectors  
to be within a bounded spherical Brillouin zone: $|{\bf k}|<\Lambda$. However, the precise value of the upper cutoff $\Lambda$ has no effect on our final  results. The fields $h({\bf x},t)$ and $\hat h({\bf x},t)$
are separated into the high and low wave vector parts: $h({\bf x},t)=h^<({\bf x},t)+h^>({\bf x},t)$, $\hat h({\bf x},t)=\hat h^<({\bf x},t)+\hat h^>({\bf x},t)$, where $h^>({\bf x},t)$ and $\hat h^>({\bf x},t)$ have support in the large wave vector  (short wavelength) 
range $\Lambda
e^{-l}<|{\bf k}|<\Lambda$, while $h^<({\bf x},t)$ and $\hat h^<({\bf x},t)$ have support in the small 
wave vector (long wavelength) range $|{\bf k}|<e^{-l}\Lambda$; $b\equiv e^{l}>1$.
We then integrate out $h^>({\bf x},t)$ and $\hat h^>({\bf x},t)$ in 
 the anhamornic coupling $\lambda$ and $\lambda_1$; as usual, this resulting perturbation theory of $h^<({\bf x},t)$,  $\hat h^<({\bf x},t)$ 
can be represented by Feynman graphs, with the order of perturbation theory 
reflected by the number of loops in the graphs we consider. We restrict ourselves here up to the one-loop order.  After this 
perturbative step, we rescale lengths, 
with ${\bf x}\rightarrow{\bf x }' e^{l}$,  which restores the UV cutoff back to 
$\Lambda$, together with rescaling of time $t\rightarrow t'e^{z l}$  (equivalently in the Fourier space, the momentum and frequency are rescaled as ${\bf k}\rightarrow {\bf k}'/b$ and $\omega \rightarrow \omega'/b^z$, respectively), where $z$ is the dynamic exponent. This is then followed
by rescaling the long wave length part of the fields that we define in the Fourier space for calculational convenience. We scale $h^<({\bf q},\omega)=\xi h(b{\bf q},b^z\omega),\,\hat h^<({\bf q},\omega)=\hat\xi \hat h(b{\bf q},b^z\omega')$. 


Next, we set $\xi \hat{\xi}= b^{d+2z}$ by demanding the term in {\em action} (\ref{act-func}) $\int d^d{\bf k} d\omega \,\hat{h} h\,\omega$ does not scale with the scaling of momenta and frequencies. Accordingly the model parameters scale as given below:
\begin{align}
 &\nu'=\nu^<b^{z-4},\, D'_h=D_h^< b^{z-d-2\chi_h},\\
 &(\lambda',\lambda_1')=(\lambda^<,\lambda_1^<) b^{z-2+\chi}.
\end{align}

\subsection{One-loop Feynman diagrams for $\nu$}

\begin{figure}[htb]
 \includegraphics[width=8cm]{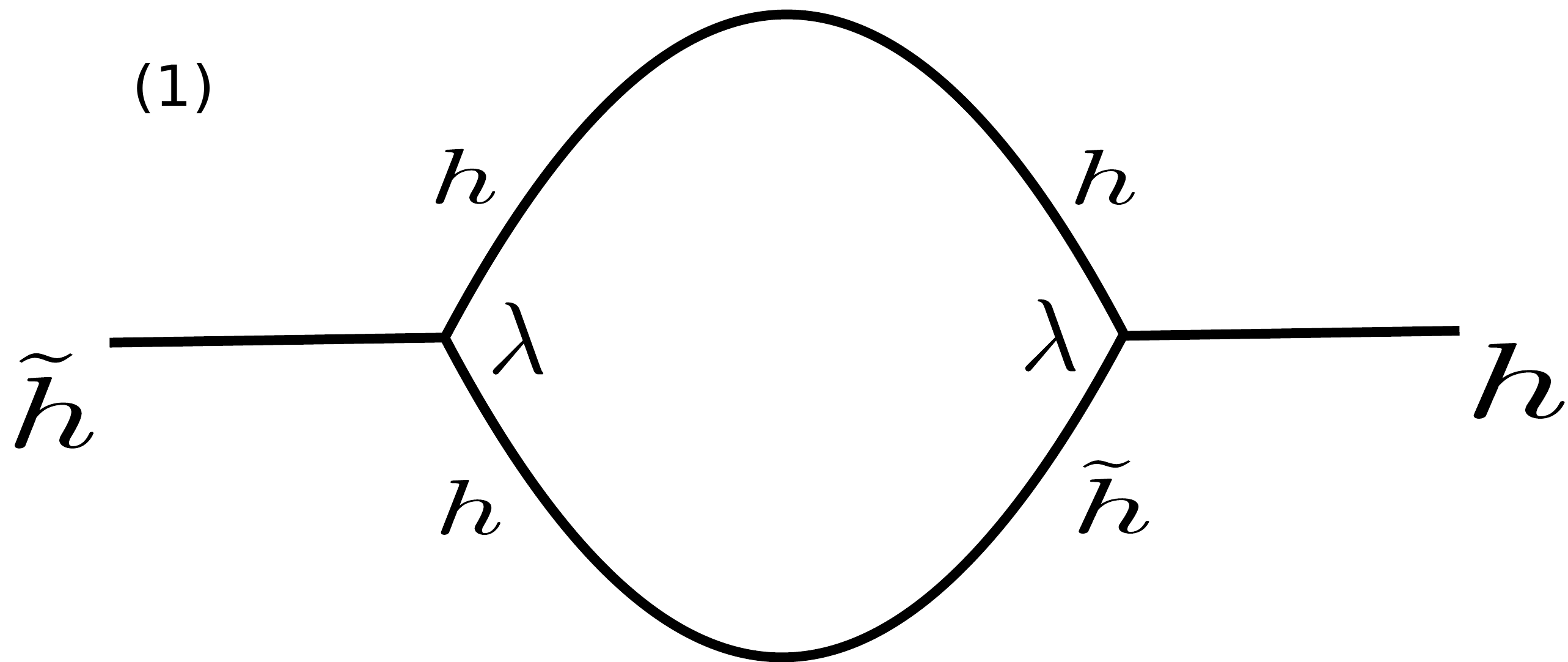}\\
 \includegraphics[width=8cm]{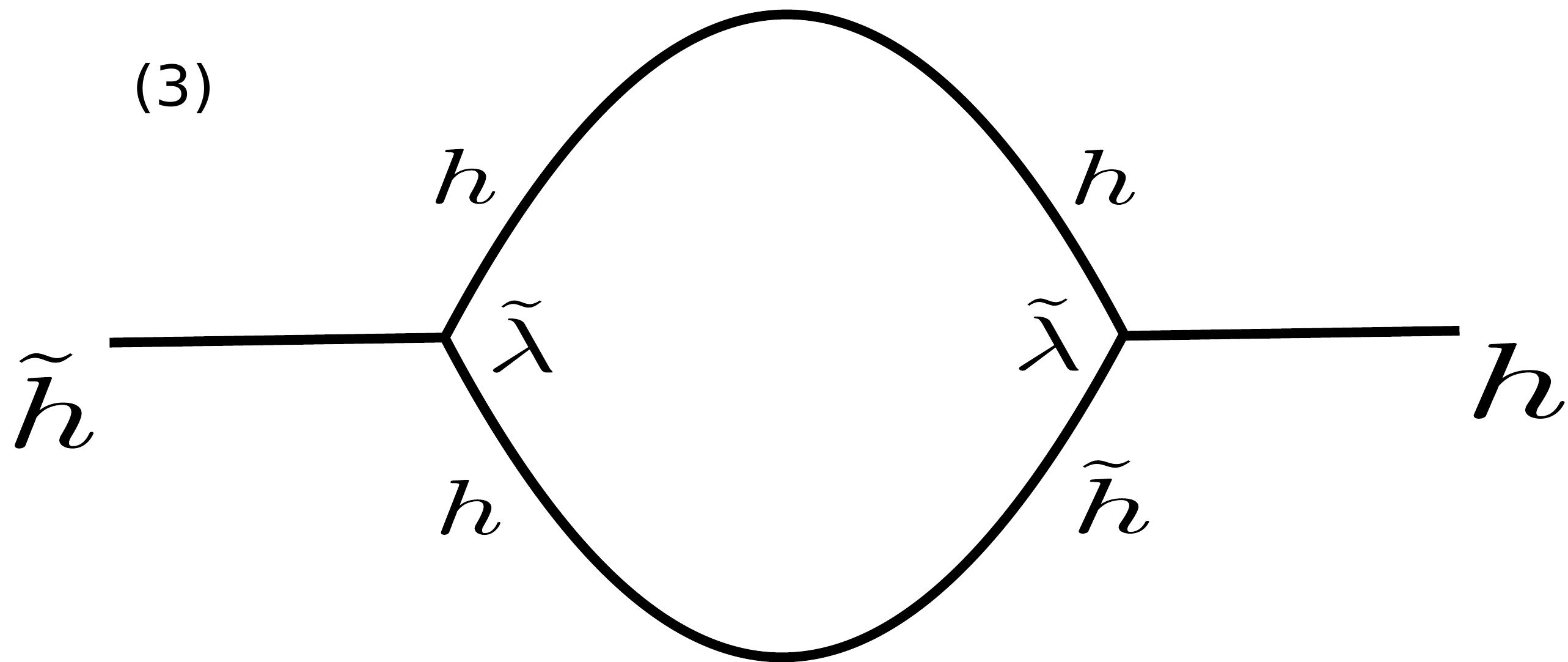}\\
 \includegraphics[width=8cm]{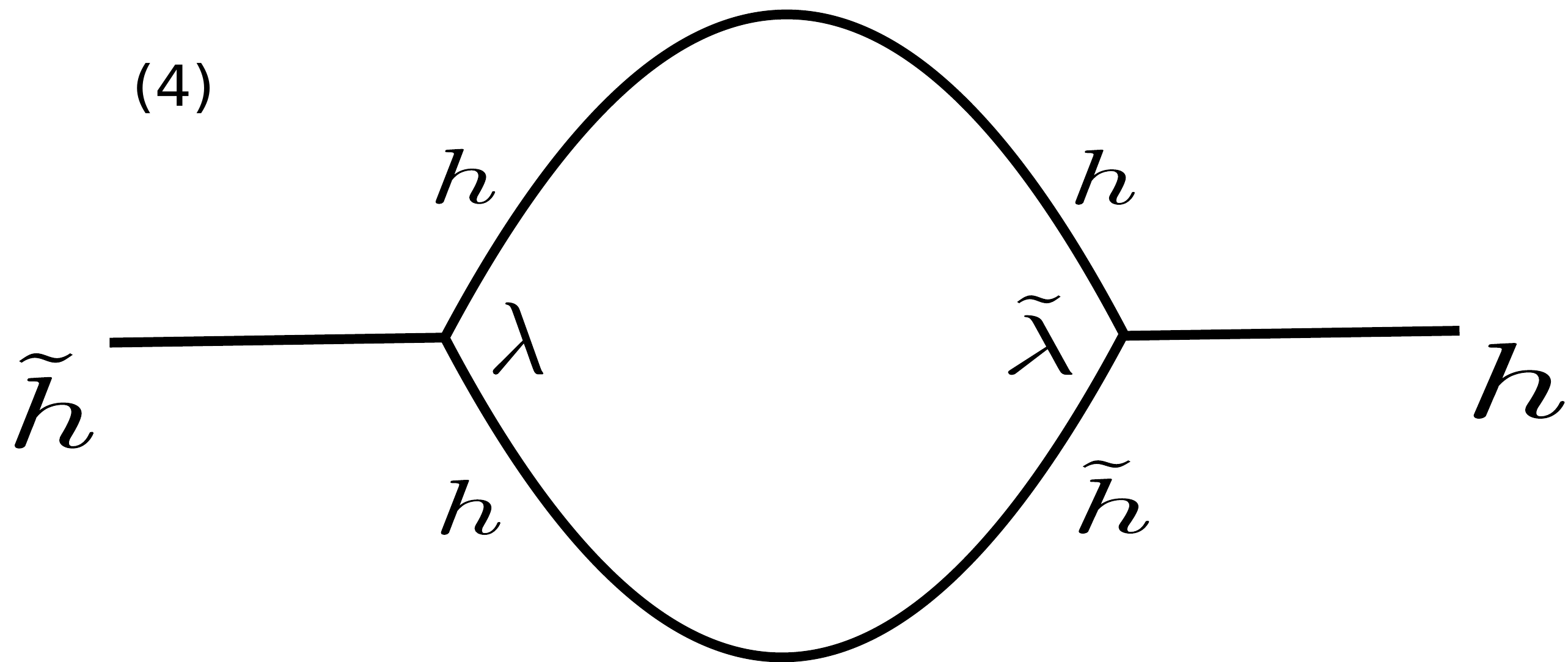}\\
 \includegraphics[width=8cm]{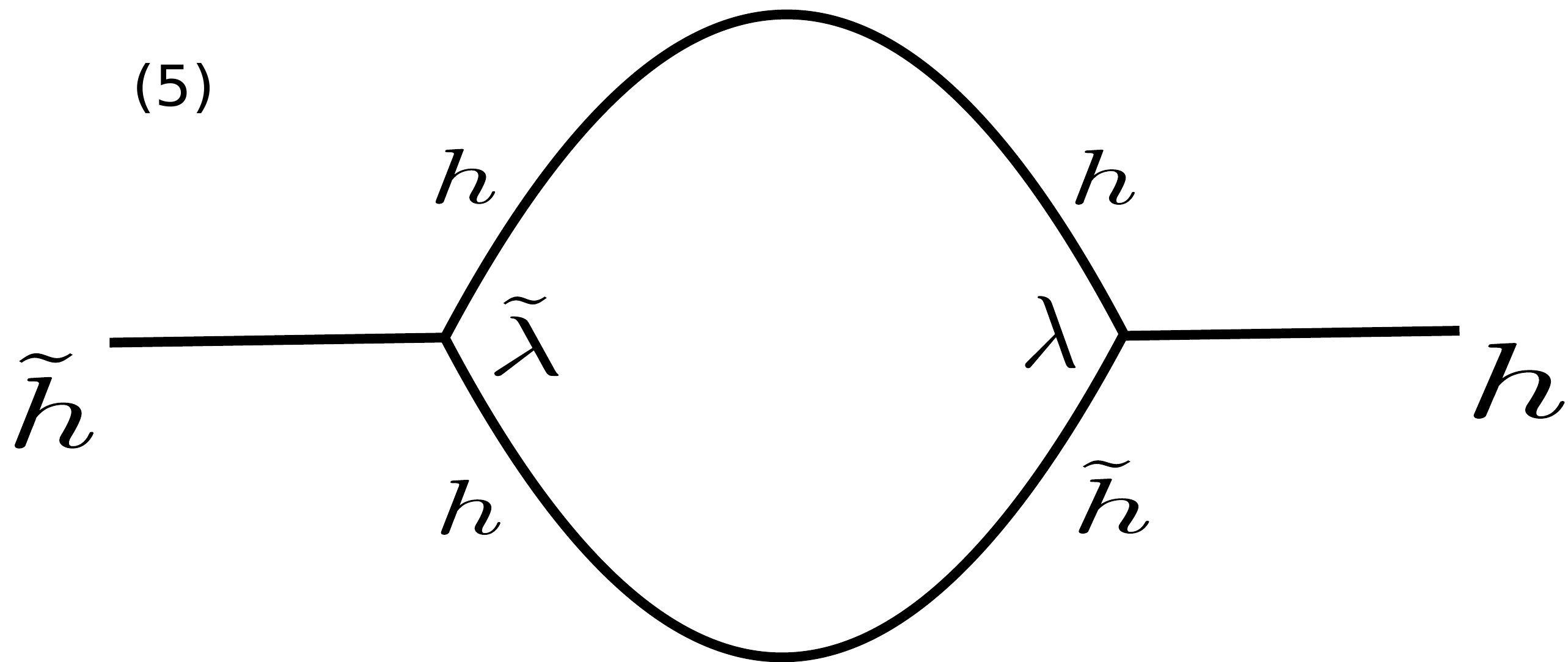}
 \caption{One-loop Feynman diagrams which contribute to the fluctuation-correction of $\nu$.}\label{nu-diag}
\end{figure}


\subsection{One-loop vertex corrections}
\begin{figure}[htb]
\includegraphics[width=8cm]{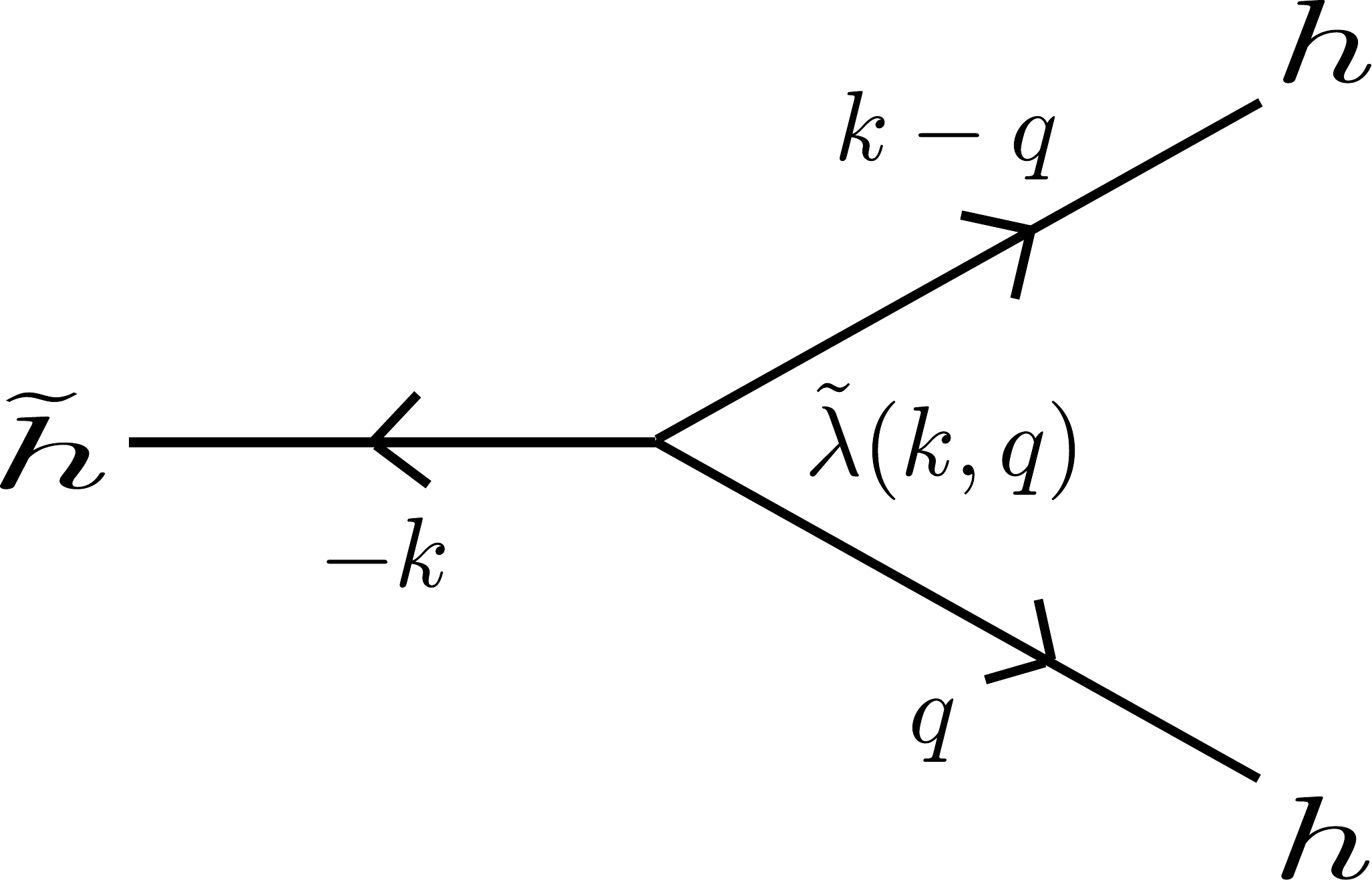}
\caption{Effective vertex $\tilde \lambda ({\bf k, \, q})=-\frac{1}{2}[q^2 {\bf k\cdot (k-q)}+ ({\bf k-q})^2 {\bf k\cdot q}]$.}\label{effver-diag}
\end{figure}


\begin{widetext}

\begin{figure}[htb]
\includegraphics[width=8cm]{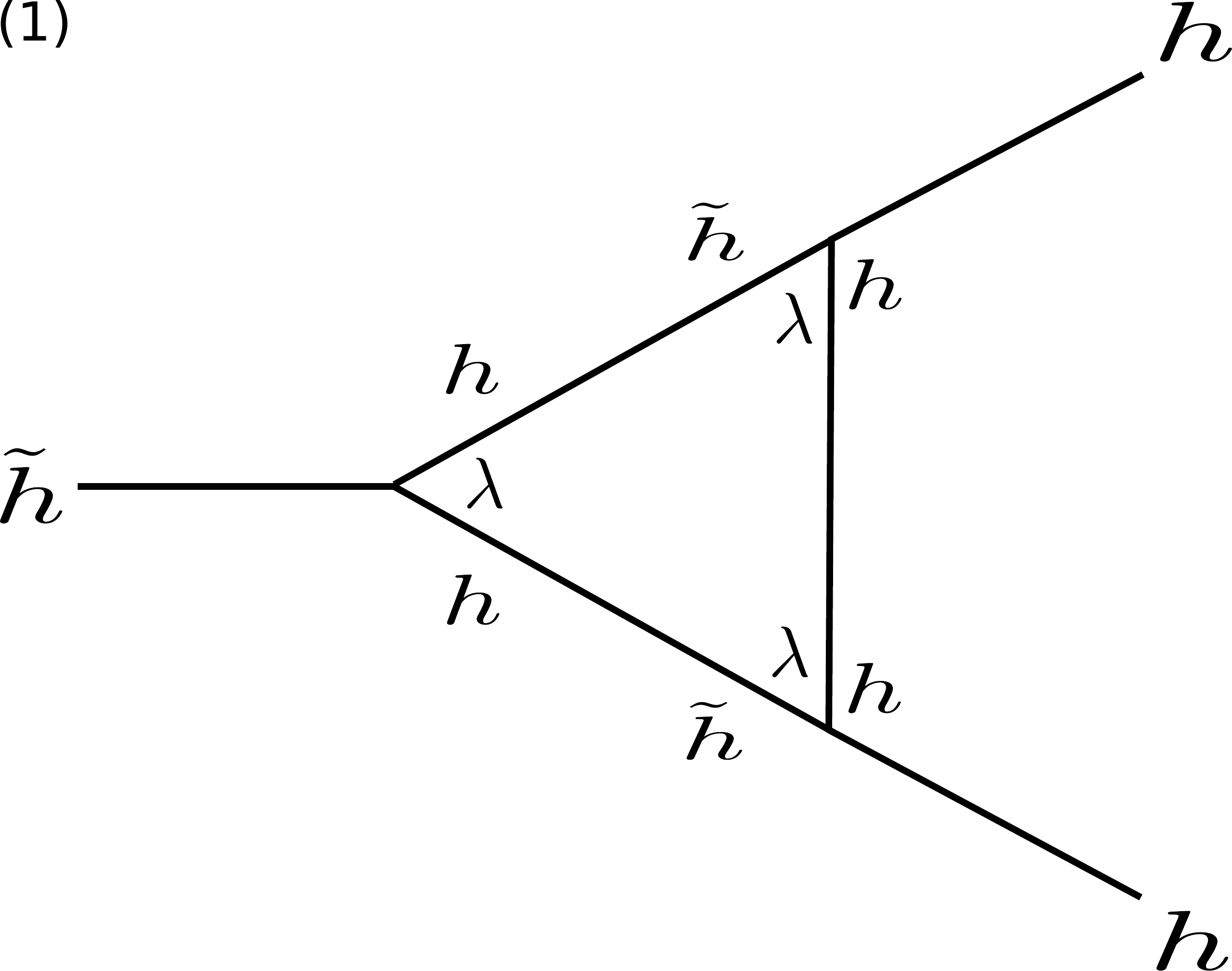}\hfill\includegraphics[width=8cm]{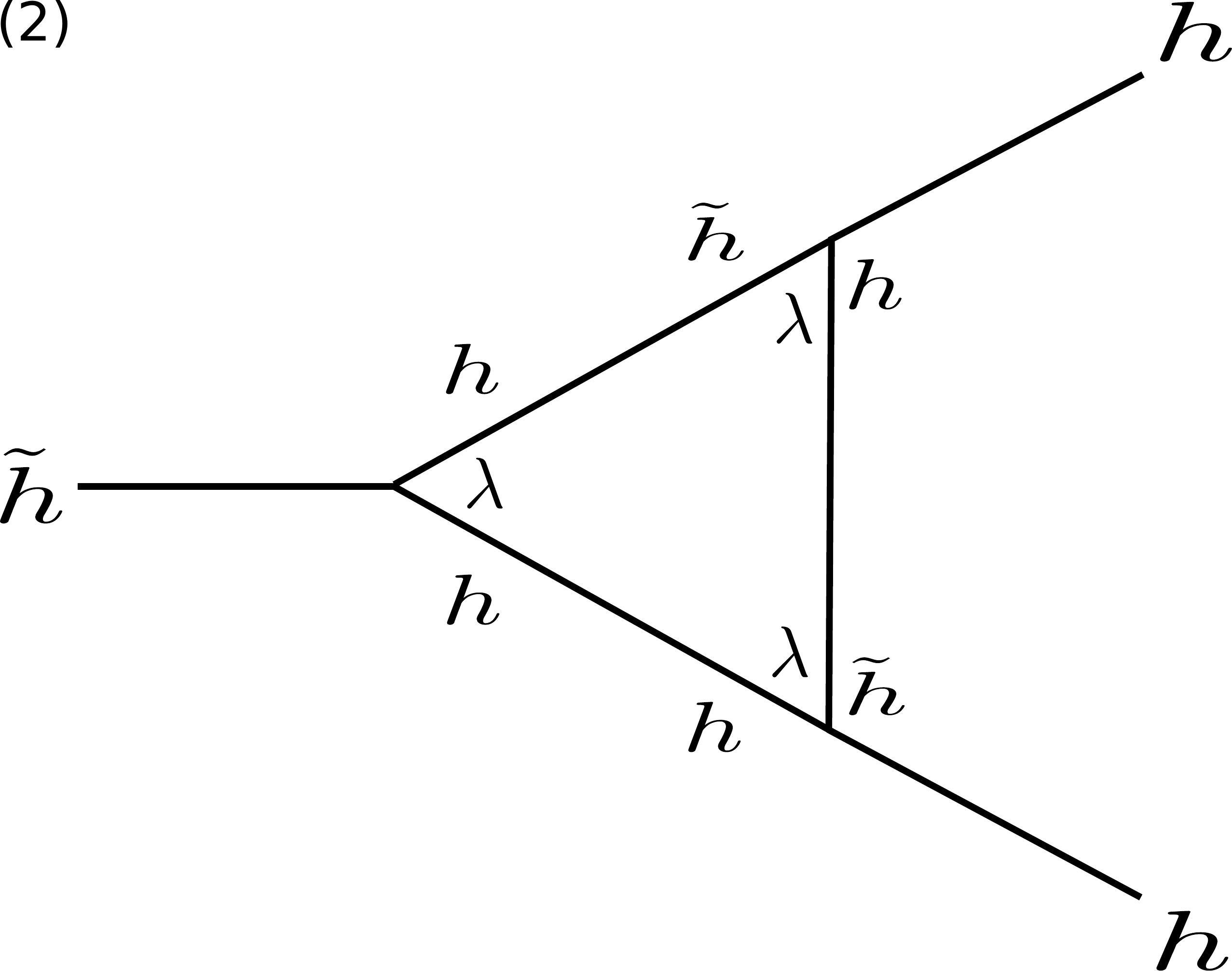}\\
 \includegraphics[width=8cm]{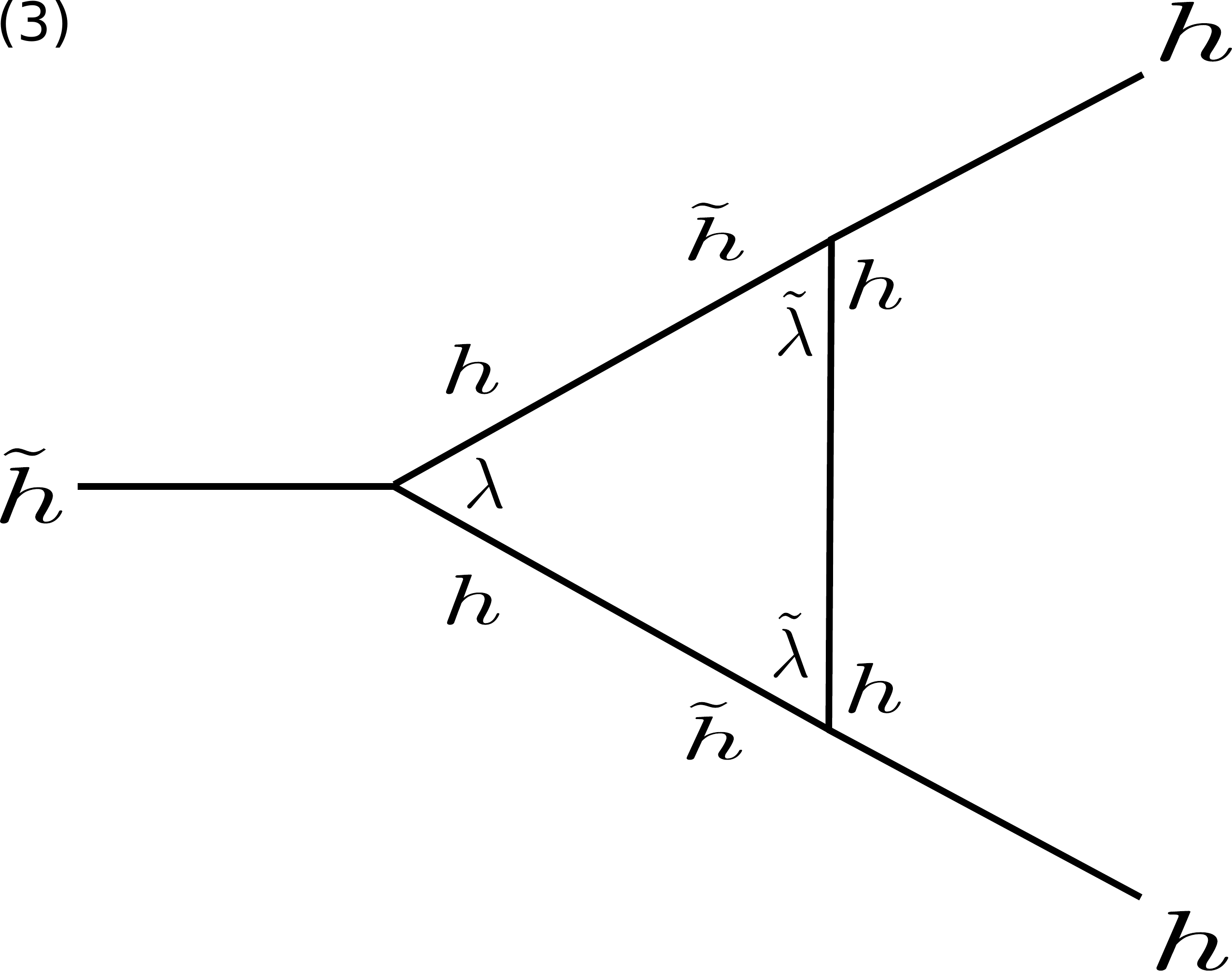}\hfill \includegraphics[width=8cm]{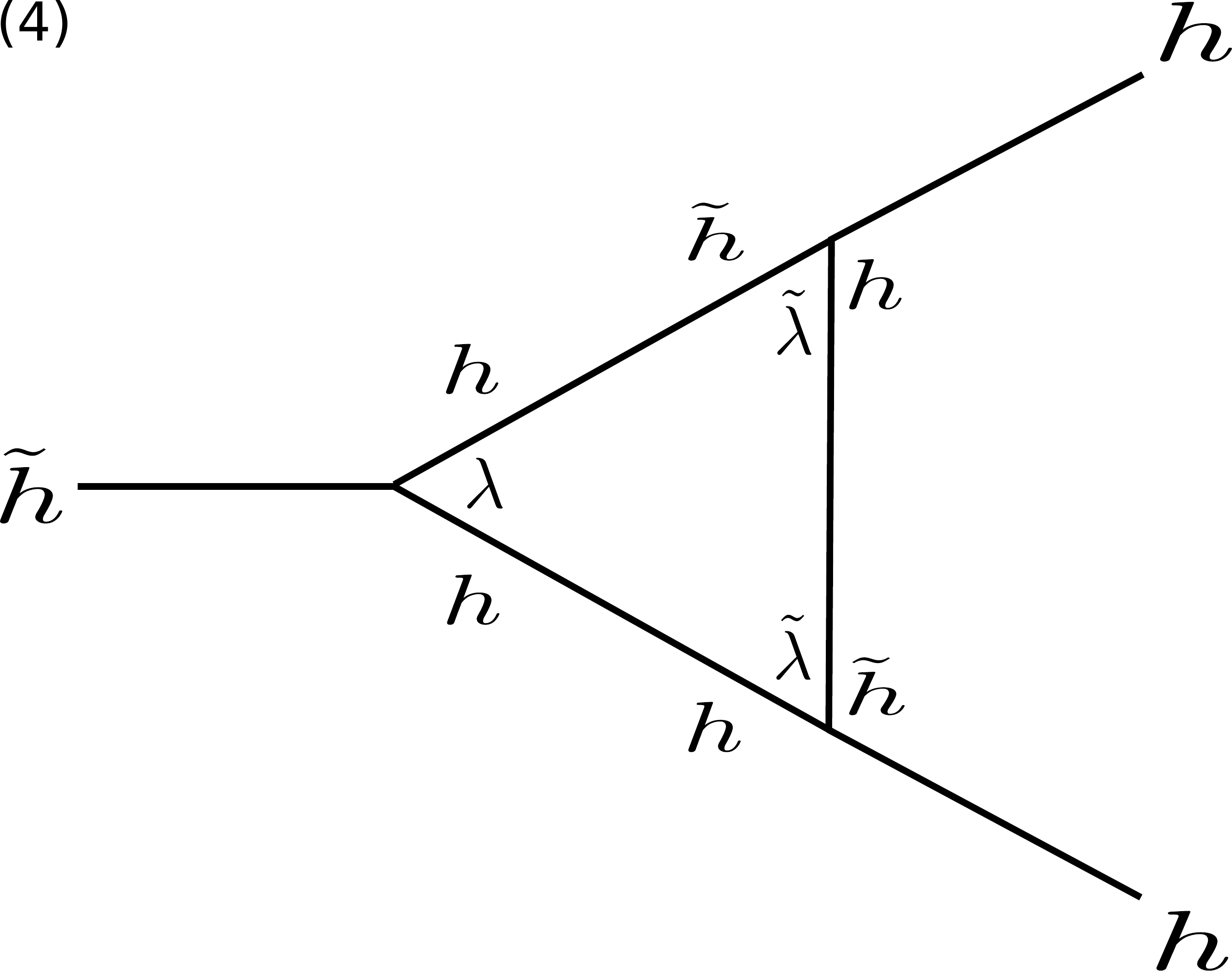}\\
 \includegraphics[width=8cm]{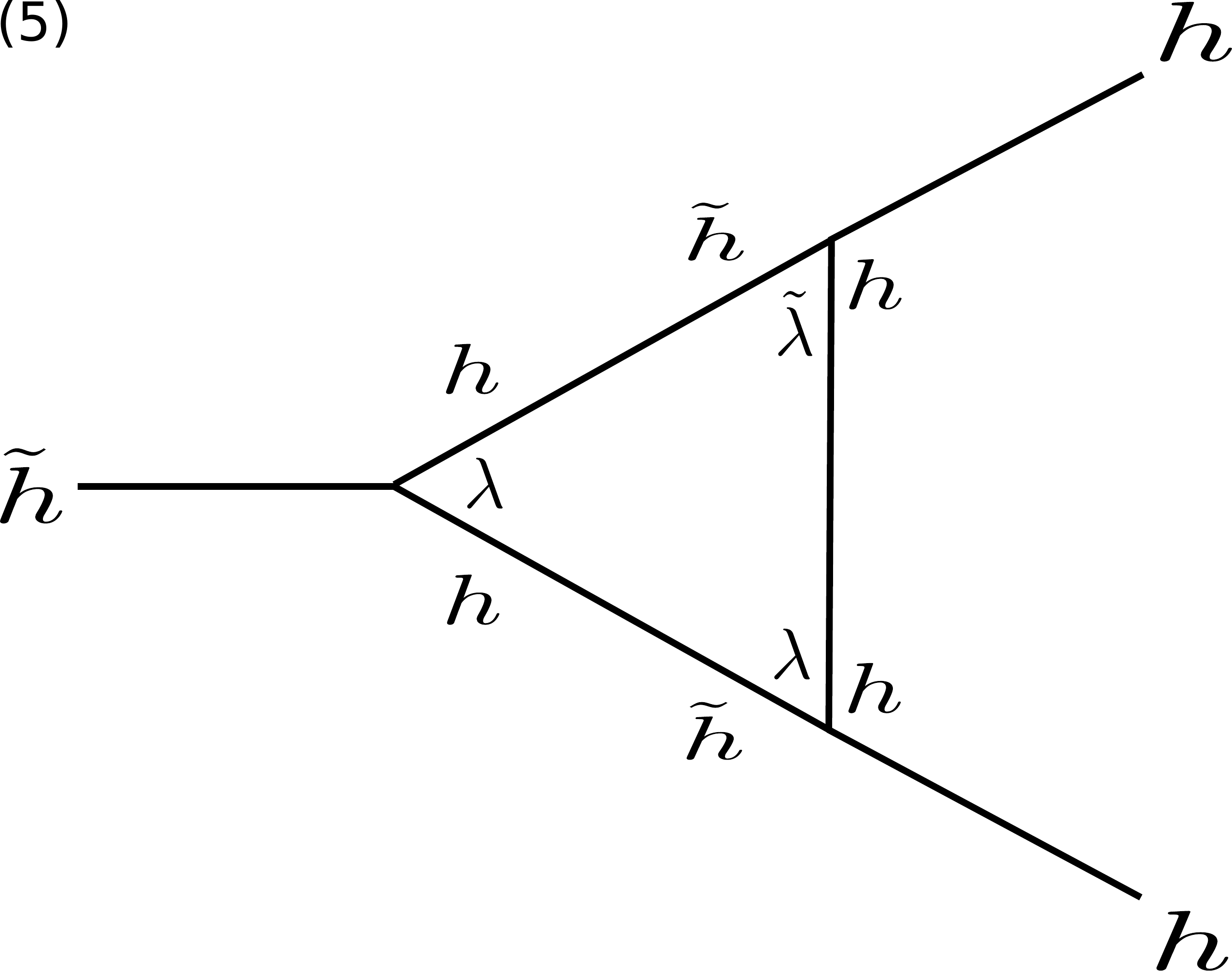}\hfill  \includegraphics[width=8cm]{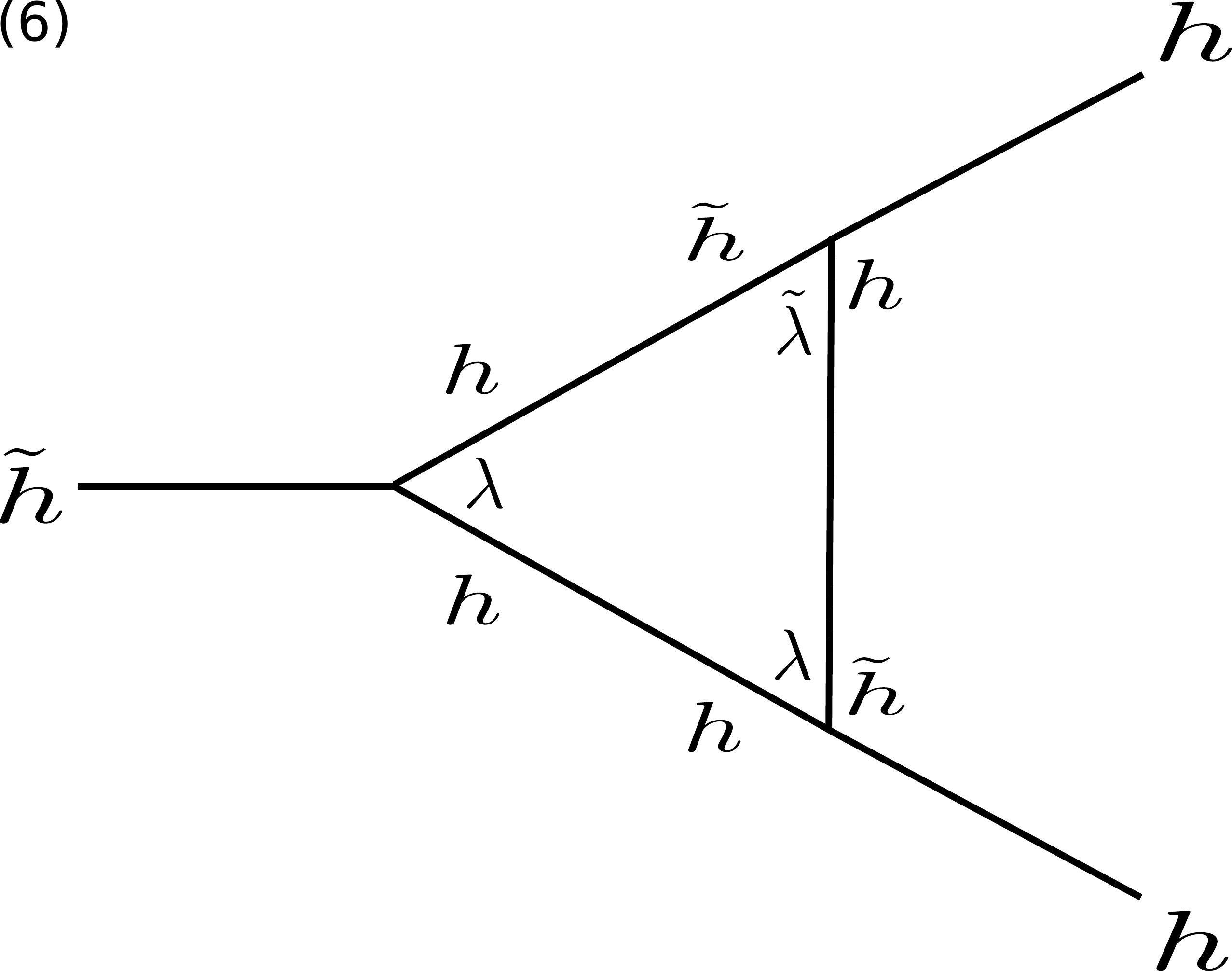} 
 \caption{One-loop Feynman diagrams which contribute to the fluctuation-corrections of $\lambda$.}\label{lamb-diag}
 \end{figure}

 Sum of these diagrams in Fig.~(\ref{lamb-diag}) vanishes.
 
 
 \newpage
 
 \begin{figure}[htb]
\includegraphics[width=8cm]{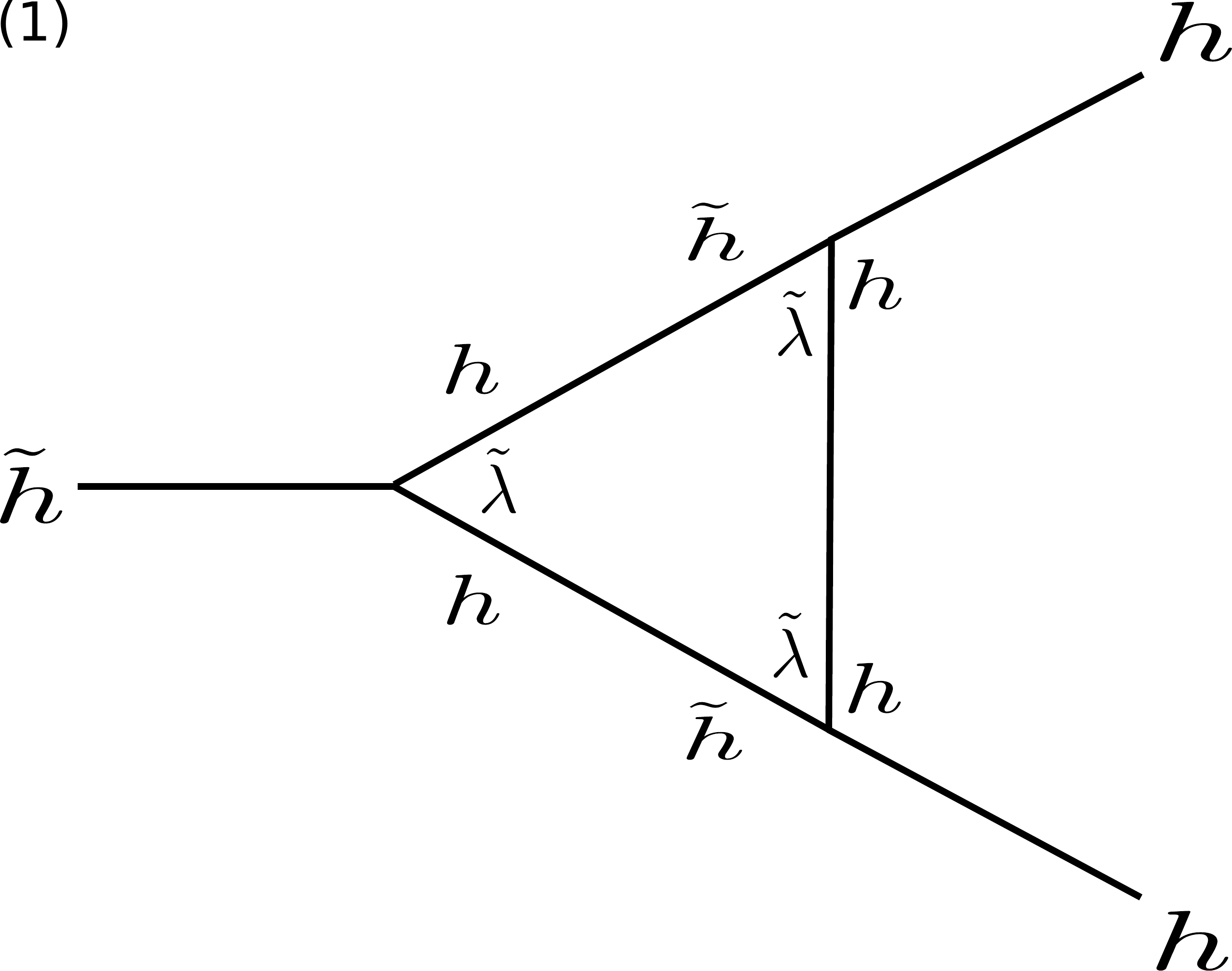}\hfill \includegraphics[width=8cm]{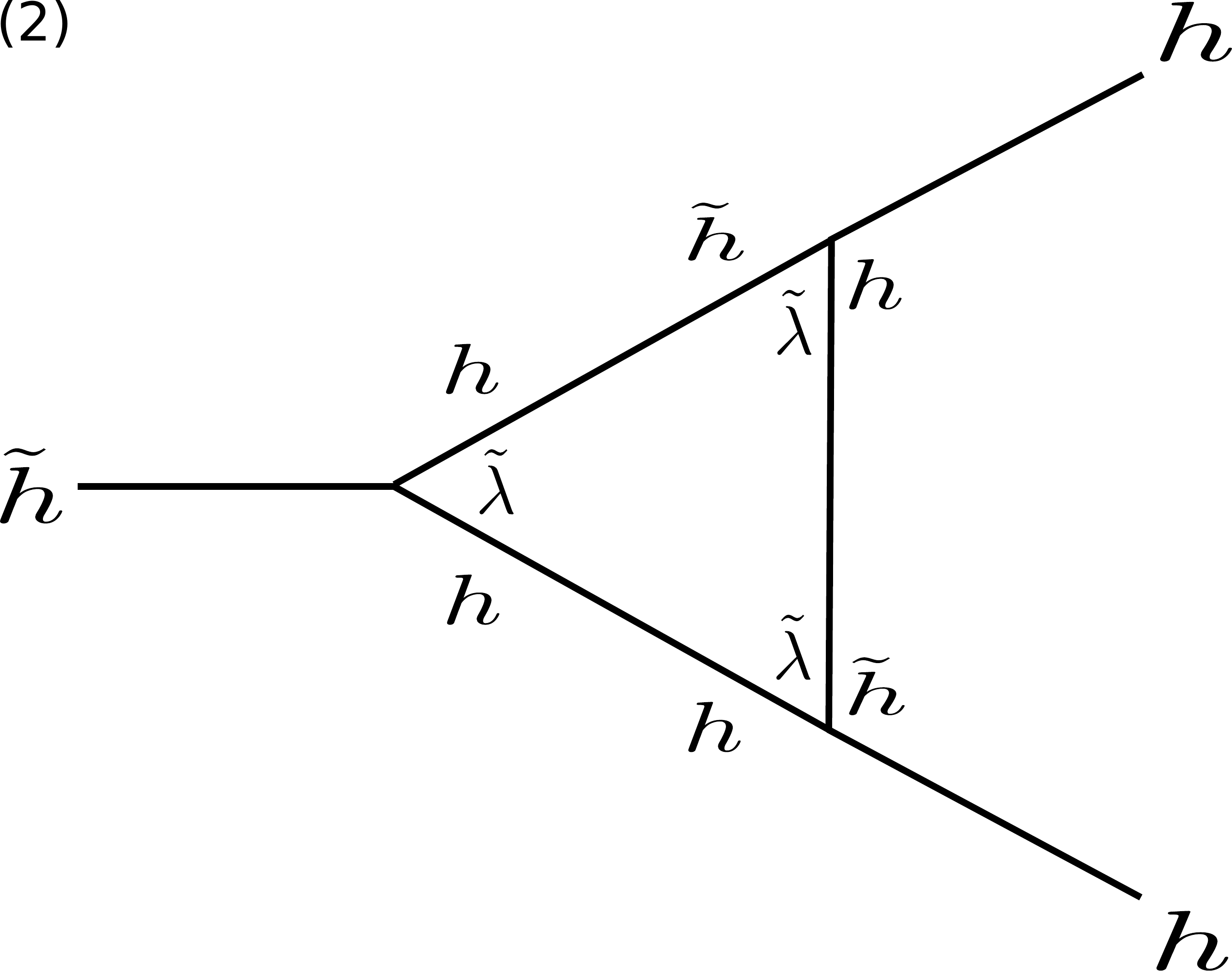}\\
 \includegraphics[width=8cm]{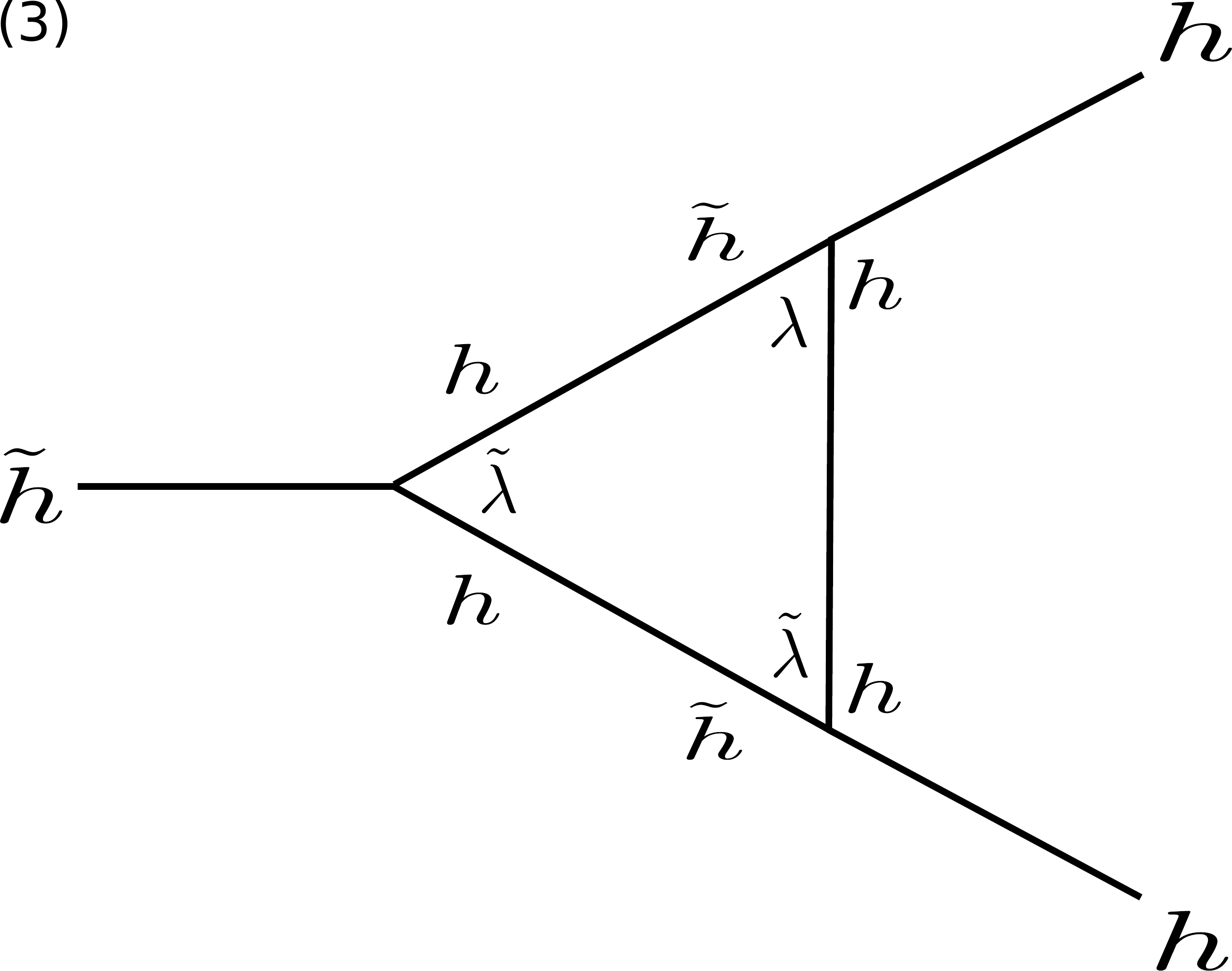}\hfill  \includegraphics[width=8cm]{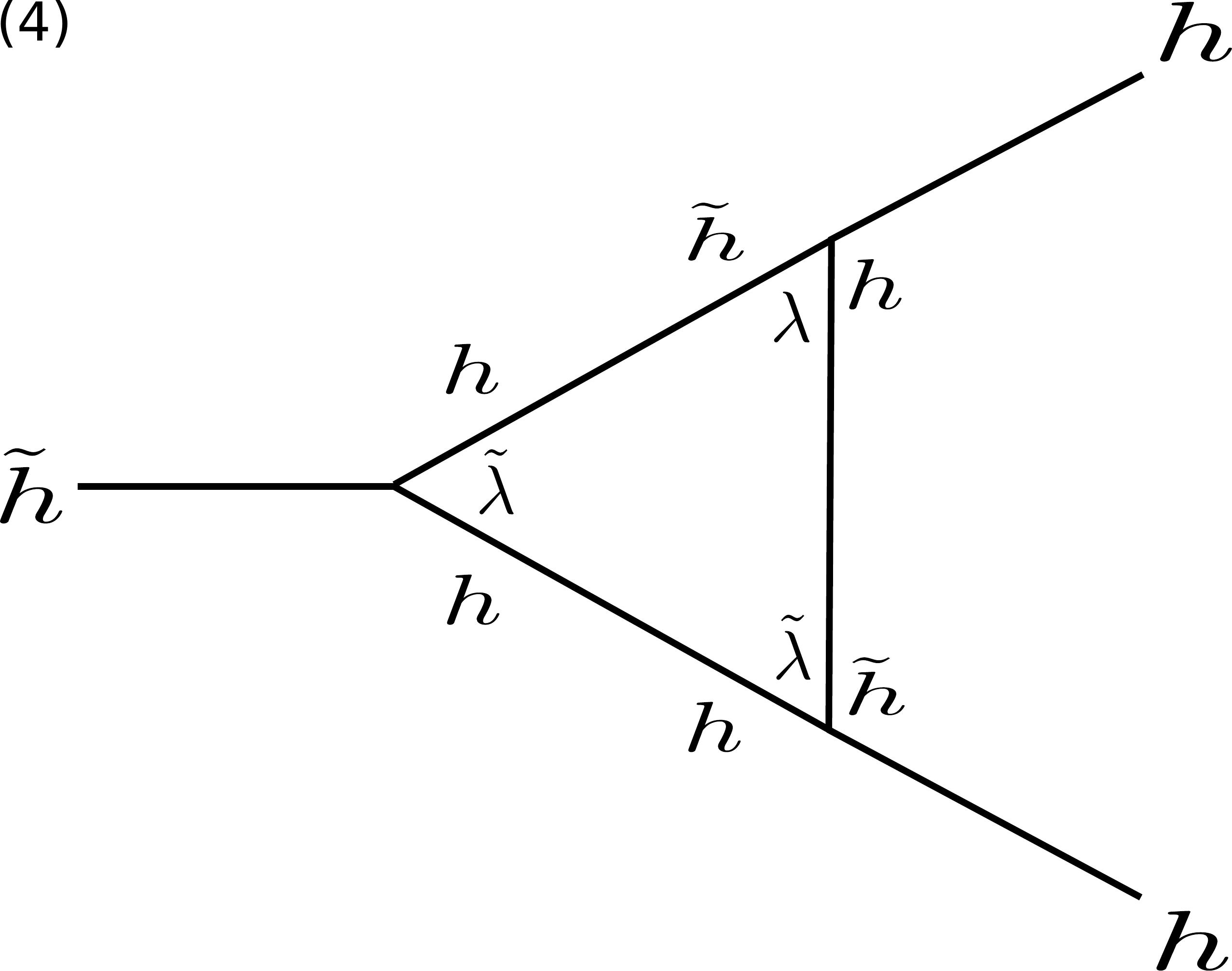}\\
 \includegraphics[width=8cm]{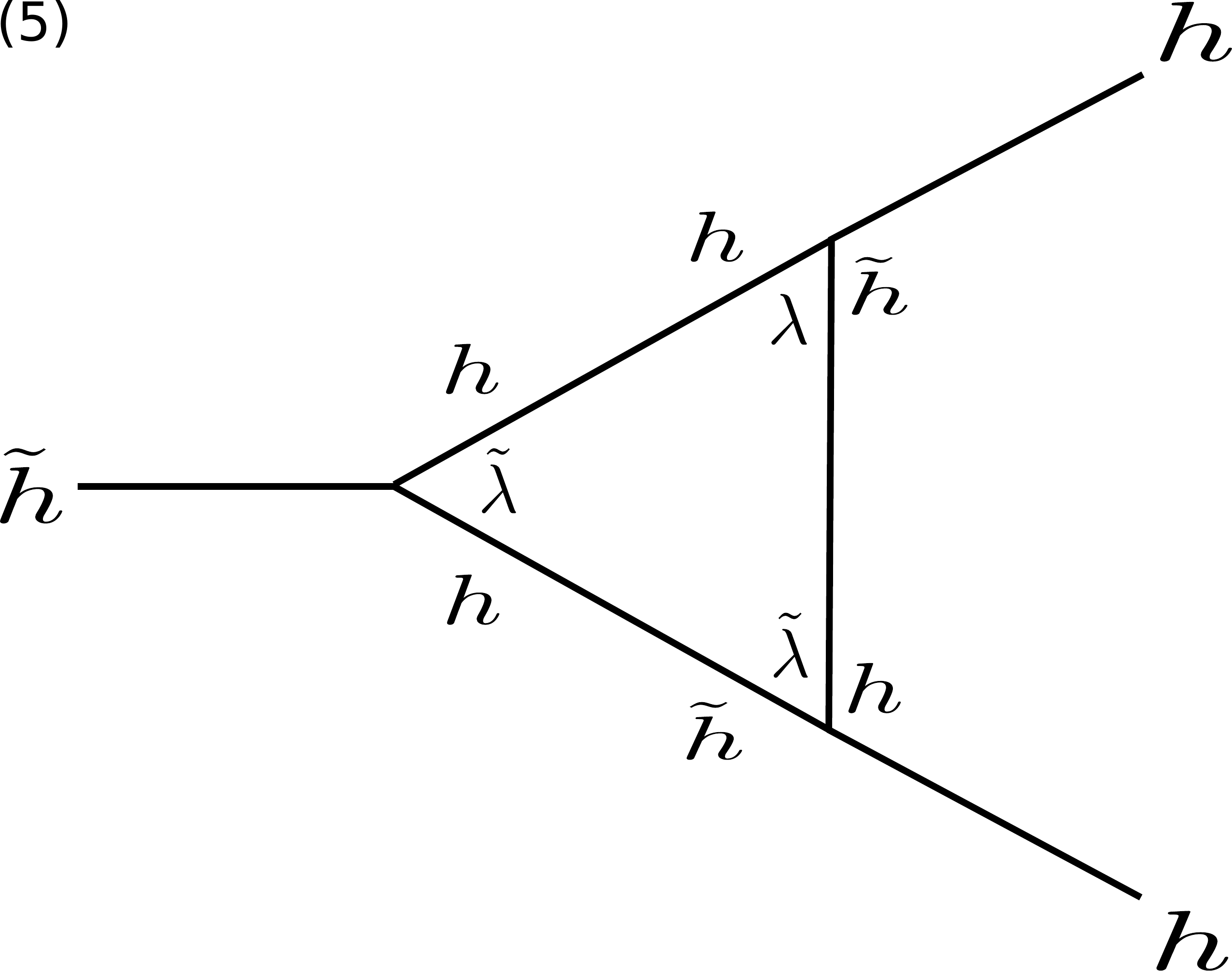}\hfill  \includegraphics[width=8cm]{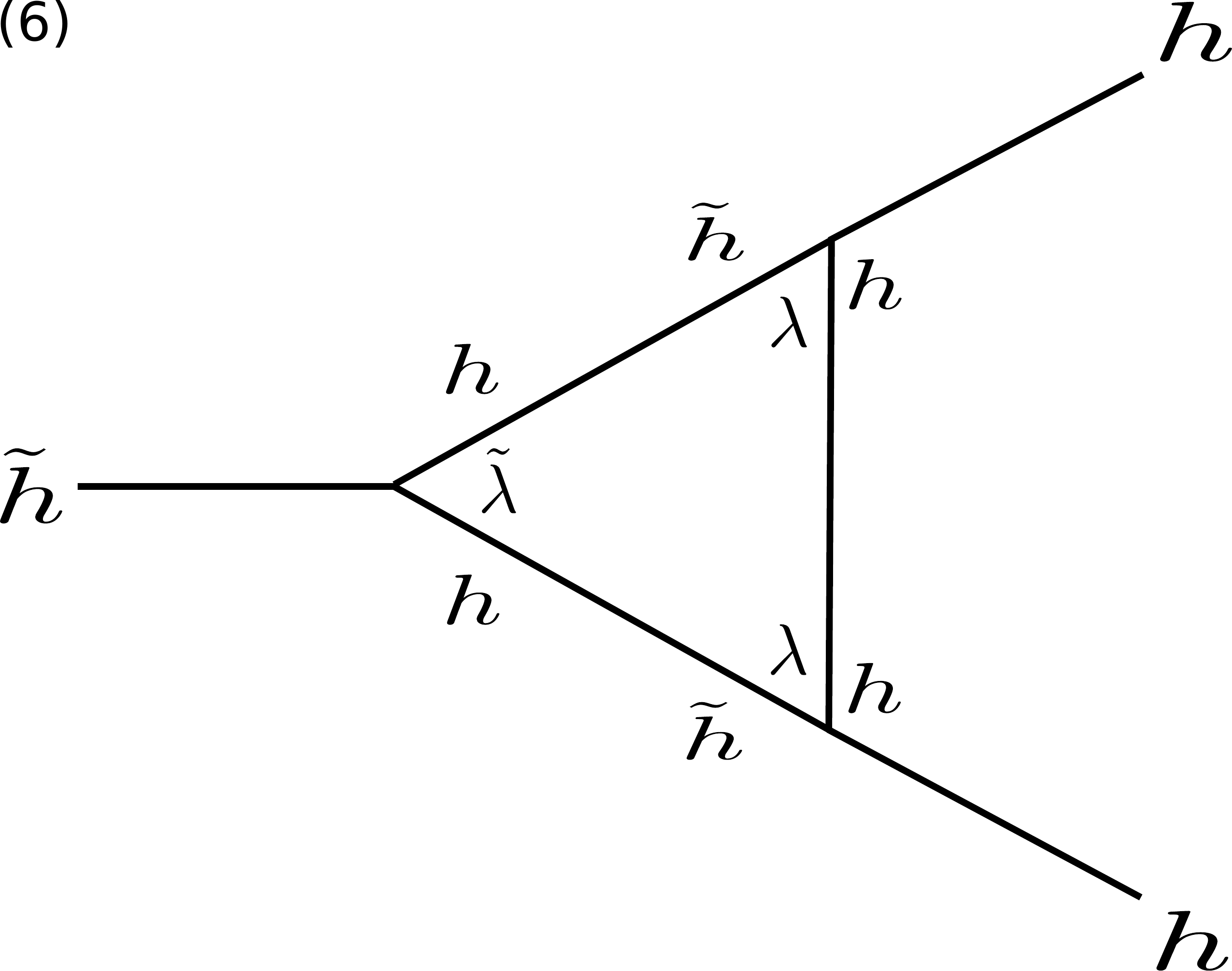}\\
 \includegraphics[width=8cm]{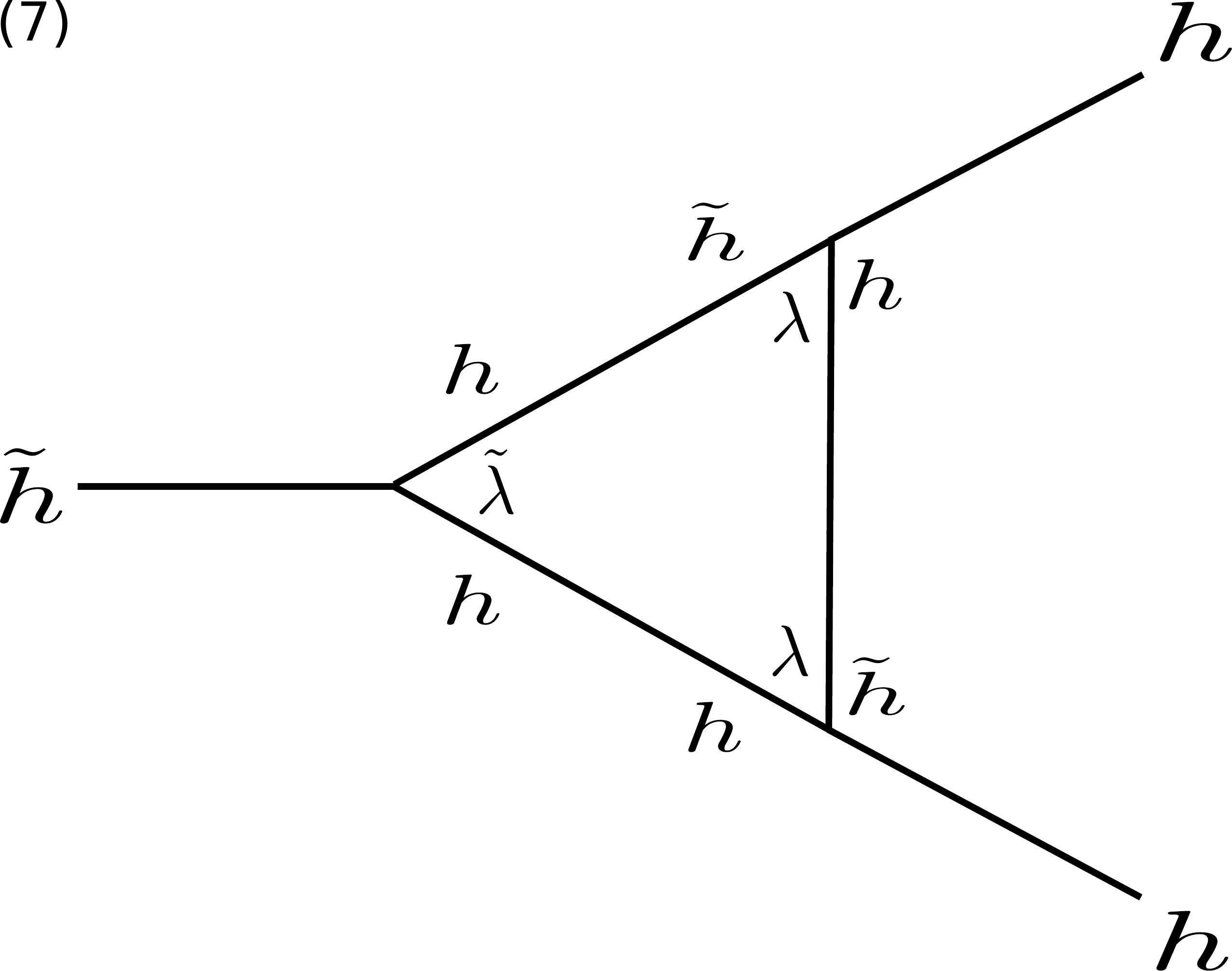}
 \caption{One-loop Feynman diagrams which contribute to the fluctuation-corrections of $\tilde\lambda$.}\label{lamb-tilde-diag}
 \end{figure}
 
 Sum of these diagrams in Fig.~(\ref{lamb-tilde-diag}) vanishes.
 
 \end{widetext}
 
 
 \section{Perturbative corrections to the model parameters}
 
 Combining all the diagrams in Fig.~(\ref{nu-diag}), and evaluating them, we obtain the fluctuation-correction diffusivity $\nu^<$:
 
 \begin{eqnarray}
  \nu^<&=& \nu \bigg[1 + \frac{{\lambda}^2D}{{\nu}^3}\frac{6-d}{4d}K_d\int_{\Lambda/b}^\Lambda dq \,q^{d-5} \cr
  &\,& -\frac{\lambda_1^2D}{{\nu}^3}K_d\int_{\Lambda/b}^\Lambda dq \,q^{d-5} \cr 
  &\,& + \frac{\lambda_1^2D}{{\nu}^3d}\left(\frac{6-d}{4} - \frac{6}{d+2} + 3\right)K_d\int_{\Lambda/b}^\Lambda dq \,q^{d-5} \cr 
  &\,& + \frac{{\lambda}{\lambda_1}D}{{\nu}^3}K_d\int_{\Lambda/b}^\Lambda dq \,q^{d-5}\bigg],\label{nu-disc}\\
   \lambda^<&=&\lambda,\\
   \lambda_1^<&=&\lambda_1,\\
   D^<&=& D.
 \end{eqnarray}

 We now rescale space, time and fields so as to raise the upper wavevector cutoff back to $\Lambda$. This gives the following differential RG flow equations for $\nu$ and $g$:
 
 \begin{eqnarray}
 \frac{d\nu}{dl}&=& \nu \bigg[z-4 + \bigg{\{}\frac{{\lambda}^2D}{{\nu}^3}\frac{6-d}{4d} -\frac{\lambda_1^2D}{{\nu}^3} \cr
 &\,& + \frac{\lambda_1^2D}{{\nu}^3d}\left(\frac{6-d}{4} - \frac{6}{d+2} + 3\right) \cr 
 &\,& + \frac{{\lambda}{\lambda_1}D}{{\nu}^3}\bigg{\}}K_d{\Lambda}^{d-4}\bigg],\\
 \frac{dg}{dl}&=&g\bigg[4-d-3 \bigg{\{}\frac{{\lambda}^2D}{{\nu}^3}\frac{6-d}{4d} -\frac{\lambda_1^2D}{{\nu}^3} \cr
 &\,& + \frac{\lambda_1^2D}{{\nu}^3d}\left(\frac{6-d}{4} - \frac{6}{d+2} + 3\right) \cr 
 &\,& + \frac{{\lambda}{\lambda_1}D}{{\nu}^3}\bigg{\}}K_d{\Lambda}^{d-4}\bigg].\label{g-flow-sm}
   \end{eqnarray}
This clearly shows that $d=4$ is the critical dimension of $g$. Near $d=4$, (\ref{g-flow-sm}) reduces to (10) of the main text.

The Feynman diagrams in Fig.~\ref{nu-diag} can be used to calculate $\nu_\text{eff}$, whence the wavevector loop integrals are done from $\Lambda$ all the way to $2\pi/L$. Clearly for $d<4$, the wavevector loop integrals depend sensitively on $L$, the system size. In fact, they diverge with $L$, a fact that opens up the intriguing possibility of crumpling with increasing $L$. In contrast, these wavevector integrals remain finite for $d>4$, even if $L\rightarrow \infty$ limit is taken.


\begin{thebibliography}{99}
\bibitem{chaikin} P. M. Chaikin and T. C. Lubensky, Principles of condensed
matter physics, Vol. 1 (Cambridge University Press, Cambridge, 2000).
\bibitem{kpz} M. Kardar, G. Parisi and Y-C. Zhang, Dynamic Scaling of Growing Interfaces,
Phys. Rev, Lett., {\bf 56}, 889 (1986).
\bibitem{stanley} A-L Barab\'asi and H. E. Stanley, Fractal concepts in surface growth (Cambridge university press, 1995).
\bibitem{kpz-rough} T. Halpin-Healy and K-C. Zhang, Kinetic roughening phenom-
ena, stochastic growth, directed polymers and all that, Phys.
Rep. 254, 215 (1995); A. Basu and E. Frey, “Novel universality classes of cou-
pled driven diffusive systems,” Phys. Rev. E 69, 015101
(2004); A. Basu and E. Frey, Scaling and universality in coupled
driven diffusive models, J. Stat. Mech.: Theory Exp.
2009, P08013 (2009).
\bibitem{ckpz-basic} T. Sun, H. Guo and M. Grant, Dynamics of driven interfaces with a conservation law, Phys. Rev. A 40, R6763 (1989).
\bibitem{mike} F. Caballero et al, Strong Coupling in Conserved Surface Roughening: A New Universality Class?, Phys. Rev. Lett.
121, 020601 (2018).
\bibitem{das-sarma} Z. -W. Lai and S. Das Sarma, Kinetic Growth with Surface Relaxation:
Continuum versus Atomistic Models, Phys. Rev. Lett. {\bf 66}, 2348 (1991). 
\bibitem{tumor} A. Br\'u, J. M. Pastor, I. Fernaud, I. Br\'u, S. Melle, and C. Berenguer, Phys. Rev. Lett. {\bf 81}, 4008 (1998); A. Br\'u, S. Albertos, J. L. Subiza, J. L. García-Asenjo, and I. Br\'u, Biophys. J. {\bf 85}, 2948 (2003).
\bibitem{tethered} M. Paczuski, M. Kardar, and D. R. Nelson, Landau Theory of
the Crumpling Transition, Phys. Rev. Lett. 60, 2638 (1988).
\bibitem{john-tethered} T. Banerjee, N. Sarkar, J. Toner, and A. Basu, Rolled up or
Crumpled: Phases of Asymmetric Tethered Membranes, Phys.
Rev. Lett. 122, 218002 (2019).
\bibitem{john-tethered1} T. Banerjee, N. Sarkar, J. Toner and A. Basu, Statistical mechanics of asymmetric tethered membranes: Spiral and crumpled phases, Phys. Rev. E 99, 053004 (2019). 
\bibitem{david-guitter} F. David and E. Guitter, Crumpling Transition in Elastic Membranes:
Renormalization Group Treatment, Europhys. Lett. {\bf 5}, 709 (1988).
\bibitem{fns} D. Forster, D. R Nelson, and M. J Stephen, 
Large-distance and long-time properties of a randomly stirred
fluid, Phys. Rev. A {\bf 16}, 732 (1977).
\bibitem{halpin} P. C. Hohenberg and B. I. Halperin, Theory of dynamic
critical phenomena, Rev. Mod. Phys. {\bf 49}, 435 (1977).
\bibitem{janssen} R Bausch, H. Janssen, and H. Wagner, Renormalized
field theory of critical dynamics, Z. Phys. B {\bf 24}, 113
(1976); U. T\"auber, Critical Dynamics (Cambridge
University Press, Cambridge, 2014).
\bibitem{sm} See Supplemental Material for details of the RG calculations including the Feynman diagrams.
\bibitem{janssen-prl} At the two-loop order, $\lambda$ does renormalise; see, H. Janssen, On Critical Exponents and the Renormalization of the Coupling Constant in Growth Models with Surface Diffusion, Phys. Rev. Lett. {\bf 78}, 1082 (1997). In the absence of a symmetry argument, presumably not only $\lambda_1$ too renormalises, $\lambda$ and $\lambda_1$ may mix under RG. This provides a technical justification to include the $\lambda_1$-term in the theory. 
\bibitem{peliti} L. Peliti and S. Leibler, Effects of Thermal Fluctuations on Systems with Small Surface Tension, Phys. Rev. Lett. 54, 1690 (1985).
\bibitem{tirtha-mem1} T. Banerjee and A. Basu, Thermal fluctuations and stiffening of symmetric heterogeneous fluid membranes, Phys. Rev. E {\bf 91}, 012119 (2015). 

\end{thebibliography}
\end{document}